\documentclass[aps,groupedaddres8s,fleqn,nofootinbib,twocolumn]{revtex4}
\usepackage{graphicx}
\usepackage{xcolor}
\usepackage{amsmath,amssymb}
\usepackage{float}
\usepackage{physics}
\usepackage{amsfonts}
\usepackage{verbatim}
\usepackage{flushend}
\usepackage{balance}
\usepackage{endnotes}
\usepackage{footnote}
\usepackage{adjustbox}
\usepackage{gensymb}
\usepackage{float}
\usepackage{epigraph}
\usepackage{xcolor}
\usepackage[utf8]{inputenc}
\usepackage{setspace}

\newcommand{\beq}{\begin{eqnarray}}
\newcommand{\eeq}{\end{eqnarray}}
\usepackage{amsmath}

\begin{document}

\title{Properties of condensed matter from fundamental physical constants}
\author{K. Trachenko$^{1}$}
\address{$^1$ School of Physical and Chemical Sciences, Queen Mary University of London, Mile End Road, London, E1 4NS, UK}

\begin{abstract}
Fundamental physical constants play a profound role in physics. For example, they govern nuclear reactions, formation of stars, nuclear synthesis and stability of biologically vital elements. These are high-energy processes discussed in particle physics, astronomy and cosmology. More recently, it was realised that fundamental physical constants extend their governing reach to low-energy processes and properties operating in condensed matter systems, often in an unexpected way. These properties are those we experience daily and can routinely measure, including viscosity, thermal conductivity, elasticity and sound. Here, we review this work. We start with the lower bound on liquid viscosity, its origin and show how to relate the bound to fundamental physical constants. The lower bound of kinematic viscosity represents the global minimum on the phase diagram. We show how this result answers the long-standing question considered by Purcell and Weisskopf, namely why viscosity never falls below a certain value. An accompanying insight is that water viscosity and water-based life are well attuned to fundamental constants, adding another higher-level layer to the anthropic principle. We then discuss viscosity minima in liquid He above and below the $\lambda$-point. We subsequently consider a very different property, thermal diffusivity, and show that it has the same minimum fixed by fundamental physical constants as viscosity. We also discuss bounds related to elastic properties, elastic moduli and their analogues in low-dimensional systems, and show how these bounds are related to the upper bound for the speed of sound. We conclude with listing ways in which the discussion of fundamental constants and bounds advance physical theories.
\end{abstract}

\maketitle

\tableofcontents

\section{Introduction}

Our search for the source of consistency and predictability of the observed physical world has led us to physical laws and related theories. These theories involve fundamental physical constants such as the Planck constant, electron mass or dimensionless combinations of these constants, pure numbers. These constants give the observed Universe its distinctive character and differentiate it from others we might imagine \cite{barrow,barrow1,carr1,carr,carrbook,finebook,cahnreview,hoganreview,adamsreview,uzanreview}.

Understanding the values of fundamental constants has a long history and is viewed as one of the grandest questions in modern science \cite{grandest}. Given that we don't know anything more fundamental \cite{weinberg}, this is probably one of the ultimate grand challenges in physics. Referring to fundamental constants as ``barcodes of ultimate reality'', Barrow proposes that these constants will one day unlock the secrets of the Universe \cite{barrow}.

Fundamental constants play a profound role in a number of processes, from governing nuclear reactions and nuclear synthesis in stars including carbon, oxygen and so on which can then form molecular structures essential to life. Theories of these processes suggest that they require a finely-tuned balance between the values of several fundamental constants. One example is the tuned balance between the masses of up and down quarks: larger up-quark mass gives the neutron world without protons and hence no atoms consisting of nuclei and electrons around them; larger down-quark mass gives the proton world without neutrons where light hydrogen atoms can form only but not heavy atoms. Our world with many heavy atoms with electronic orbitals which endow complex chemistry would disappear with only a few per cent fractional change in the mass difference of the two quarks \cite{hoganbook,hoganreview,adamsreview}.

Another commonly discussed example is the Hoyle's prediction of the energy level of carbon nucleus of about 7.65 MeV. This resonance level is required in order to explain carbon abundance and in particular the synthesis of carbon from fusing three alpha particles in stars \cite{barrow,barrow1,carrbook,finebook}. Following the Hoyle's prediction, the required energy level was experimentally confirmed. This carbon resonance-level coincidence is considered striking. A related important effect is a slightly lower resonance level in oxygen, which enables carbon to survive further resonant reactions. This finely balanced sequence of coincidences enables carbon-based life. In this process, production of carbon and oxygen importantly depends on their nuclear energy levels which, in turn, depend on the fine structure constant $\alpha=\frac{e^2}{\hbar c}\approx\frac{1}{137}$ and strong nuclear force constant. A small change of these constants (more than 0.4\% and 4\% for the nuclear and fine structure constant) results in almost no carbon or oxygen produced in stars \cite{barrow,carrbook,finebook}. $\alpha$ and the proton-to-electron mass ratio $\beta=\frac{m_e}{m_p}\approx\frac{1}{1836}$ play a role in making the centres of stars hot enough to initiate nuclear reactions, and unless $\alpha$ and $\beta$ satisfy a certain relation, there would be heavy nuclei produced in stars. There are other examples of what would happen as a result of altering fundamental constants, all showing that there is a fairly narrow ``habitable zone'' in the parameter space ($\alpha$,$\beta$) (see, however, Ref. \cite{adamsreview}). In this zone,  matter can remain stable long enough for stars to evolve and produce essential biochemical elements including carbon, planets can form and life-supporting molecular structures can emerge \cite{barrow,barrow1,carrbook,finebook}. For this reason, the observed fundamental constants are called ``bio-friendly'' or ``biophilic \cite{barrow,adamsreview}.

The discussion of the role of fundamental constants was mostly limited to high-energy processes including particle physics, astronomy and cosmoslogy. More recently, it has been realised that the fundamental constants extend their governing reach to the properties of condensed matter phases and at energy much lower than the high-energy physics. Many of these properties are those we experience daily and can routinely measure, including viscosity, thermal conductivity, elasticity and sound. Although these are all familiar properties, their numerical values remain hard to predict on the basis of an analytical theory because they are strongly depend on the system and external parameters. This is contrast to a class of universal properties such as, for example, the Dulong-Petit result for the specific heat.

One frequent way in which fundamental physical constants affect system properties is that they impose a {\it bound} on a property. We will show that a number of important physical properties have lower or upper bounds in a sense that they do not fall below or exceed certain values. Understanding the origin of these bounds has enthralled physicists, including those interested in collective dynamics and systems where many interacting agents operate. Apart from the interest in the values and origins of the bounds themselves, there is another important reason why bounds are interesting: finding and understanding these
bounds often means that we enhance our grasp of or clarify the underlying physics or property in question.

The main aim of this review is to summarise and synthesise earlier and more recent results related to condensed matter properties in terms of fundamental physical constants. In the process, we will see that comparing the observed properties to their fundamental bounds reveals important insights not just about the bounds themselves but also about the essential physical processes at operation as well as theories of those processes. This includes understanding different dynamical regimes of the system and predicting its behavior in future experiments.

This reviews is organised as follows. In Chapter \ref{minimalv}, we discuss the lower bound on liquid viscosity, its origin and show how to relate this bound to fundamental constants. We show how this result answers the long-standing question posed by Purcell and considered by Weisskopf, namely why viscosity never falls below a certain value. This has the implications for water viscosity and life which appears to be well attuned to the degree of quantumness of the physical world and other fundamental constants, providing another (biochemical) layer to the discussion of the anthropic principle. We will note that the viscosity minimum is interestingly close to that in a very different system, the quark-gluon plasma. We also discuss viscosity minima in liquid He above and below the $\lambda$-point.

In Chapter \ref{thermal}, we consider a very different property, thermal conductivity, and show that, similarly to viscosity, it has a minimum fixed by fundamental constants. Whereas thermal diffusivity minimum gives a minimum on the phase diagram except in the vicinity of the critical point, the minimum of kinematic viscosity is a global minimum on the entire phase diagram as discussed in Chapter \ref{minimaphase}.

In Chapter \ref{elastic}, we review the bounds related to elastic moduli and their analogues in low-dimensional systems. This will lead us to the last Chapter \ref{sound} where we discuss the upper bound on the speed of sound in condensed matter phases. Our review includes fairly recent results including our own, and we raise interesting open questions in this and related fields.

In the last Chapter \ref{summary}, we conclude with listing ways in which the discussion of fundamental constants and bounds advance physical theories. This includes insights about essential physical processes at operation, understanding different dynamical regimes, predicting future experiments as well as understanding characteristic values of condensed matter properties. The realisation that water-based life forms are well attuned to fundamental constants raises far-reaching questions related to our place in the Universe, e.g. what values of fundamental constants make water-based life possible and how well-tuned these constants need to be to remain bio-friendly at the biochemical level.

\section{Minimal viscosity}
\label{minimalv}

\subsection{The liquid problem}

Our first case study involves viscosity and its minima. We show that the minimal value of liquid viscosity turns out to be nearly universal and set by the fundamental physical constants. Here we encounter the first example of what we mentioned in the Introduction: fundamental constants impose bounds on condensed matter properties.

That viscosity minima of all liquids are universal is remarkable and unexpected for two reasons. First, the universal result applies to a variety of liquid systems, with different structure, chemistry and intermolecular interactions. The second reason is that problems involved in the liquid theory are  fundamental. To appreciate the second point, we briefly review it below.

Properties of real liquids have proved to be particularly hard to understand and calculate theoretically. Common liquid models are inapplicable to understanding the energy and heat capacity of real liquids. These models include notable workhorses of liquid physics: the widely discussed Van der Waals mode and the hard-spheres model \cite{hansen2,ziman,march,parisihard}. Both models give the specific heat $c_v=\frac{3}{2}k_{\rm B}$ \cite{landaustat,wallacecv}, the ideal-gas value, in contrast to experiments showing liquid $c_v=3k_{\rm B}$ close to melting \cite{wallacecv,wallacebook,proctor2}. These models were also used as reference states to calculate the energy \eqref{enint} by expanding interactions into repulsive and attractive parts (see, e.g., Refs. \cite{barkerhenderson,wca1,wca2,zwanzig,rosentar}). These parts understandably play different roles at high and low density, however this method faces the problem that interactions and expansion coefficients are strongly system-dependent and so are the final results, precluding a general theory. This is part of a more general problem stated by Landau, Lifhitz and Pitaevskii and discussed below.

As stated by Landau, Lifshitz and Pitaevskii (LLP), the absence of a small parameter due to the combination of strong interactions and the absence of small oscillations disallows a possibility of calculating liquid thermodynamic properties in general form \cite{landaustat,Pitaevskii}. Lets consider the calculation of liquid energy as

\begin{equation}
E=\frac{3}{2}NT+\frac{n}{2}\int g(r)u(r)dV
\label{enint}
\end{equation}

\noindent where $n$ is concentration, $g(r)$ is the pair distribution function, $u(r)$ is the interaction potential, interactions and correlations are assumed to be pairwise. Here and below, $k_{\rm B}=1$.

Since the interaction $u(r)$ in liquids is both strong and system-specific, $E$ in Eq. \eqref{enint} is strongly system-dependent. For this reason, no generally applicable theory of liquids is considered possible as discussed by LLP \cite{landaustat,Pitaevskii}. An additional difficulty is that interatomic interactions and correlation functions are not available apart from fairly simple model liquids such as Lennard-Jones systems and can be generally complex involving many-body, long-range and hydrogen-bonded interactions. The interactions and correlation functions can be simulated quantum-mechanically or obtained from experiments. This is a hard task which, if achievable, reduces the predictive power of a theory. Even when $g(r)$ and $u(r)$ are available in simple cases, the calculation involving Eq. \eqref{enint} is not enough: one still needs to develop a physical model explaining experimental temperature dependence of energy and heat capacity of real liquids \cite{ropp}. Such a general model based on interactions and correlation functions (exemplified by Eq. \eqref{enint}) has not emerged.

In solids, the above issues do not emerge because the solid state theory is based on collective excitations, phonons. This theory is predictive, physically transparent and generally applicable to all solids. There is no need to explicitly consider structure and interactions in order to understand basic thermodynamic properties of solids. Most important results such as universal temperature dependence of energy and heat capacity readily come out in the phonon approach to solids \cite{landaustat}. The simplifying small parameter in solids are small phonon displacements from equilibrium, but this seemingly does not apply to liquids because liquids do not have stable equilibrium points that can be used to sustain these small phonon displacements. Weakness of interactions used in the theory of gases does not apply to liquids either because interactions in liquids are as strong as in solids. This constitutes the no small parameter problem outlined by LLP \cite{landaustat,Pitaevskii}.

It is therefore interesting to observe that earlier liquid theories and the solid state theory diverged at the point of a fundamental approach. Early liquid theories  \cite{kirkwoodbook,borngreen,zwanzig,barkerhenderson,wca1} considered that the goal of the statistical theory of liquids is to provide a relation between liquid thermodynamics and liquid structure and intermolecular interactions such as $g(r)$ and $u(r)$ in Eq. \eqref{enint}. Working towards this goal involved developing the analytical models for liquid structure and interactions, which has become the essence of earlier liquid theories \cite{egelstaff,faber,march,marchtosi,tabor,faber1,balucani,hansen2,hansen1}. The solid state theory, on the other hand, does not aim to predict the solid structure and its characteristics such as $g(r)$. For a given chemical composition, the structure can be predicted in quantum-mechanical calculations \cite{pickard, pickard1} but not by a purely theoretical approach. Instead, the structure is often an input to theory. Similarly, the solid state theory does not aim to predict interatomic interactions. Some simple models of these interactions play a useful role in the solid state theory, however the variety of interactions (ionic, covalent and their combinations, metallic, dispersion, hydrogen-bond interactions and so on) belongs to the realm of computational physics or chemistry rather than pure theory.

Although the approach to the liquid theory diverged from the solid state theory in its fundamental perspective, there were notable exceptions. Sommerfeld \cite{somm} and Brillouin \cite{br1,br2,br3,brillouin} considered that the liquid energy and thermodynamic properties are fundamentally related to phonons as in solids and discussed liquid properties on the basis of a modified Debye theory of solids. The first Sommerfeld paper discussing this was published only 1 year after the Debye theory of solids \cite{debyepaper} and 6 years after the Einstein's paper ``Planck’s theory of radiation and the theory of the specific heat'' in 1907 \cite{einstein}. Apart from isolated attempts \cite{wannier,faber,wallacebook}, this line of enquiry has stalled in the years that followed, and liquid theories based on structure and interactions were pursued instead. Whereas the Debye and Einstein theories have become part of nearly every textbook where solids and phonons are mentioned, a theory of liquid thermodynamics has remained unworkable for about a century that followed. One potential reason for this is that, differently from solids, the nature of collective excitations in liquids remained unclear for a long time.

As a result of these issues, theoretical calculation and understanding energy and heat capacity of real classical liquids (both its values and temperature dependence) has remained a long-standing problem in both research and undergraduate teaching \cite{granato,chen-review,prescod}.

The problems involved in liquid theory started to lift fairly recently and involved several steps. The first step involved the consideration of microscopic dynamics of liquid particles provided by the Frenkel theory \cite{frenkel}: differently from solids where particle dynamics is purely oscillatory and gases where dynamics is purely diffusive/ballistic, particle dynamics in liquids is mixed and combines oscillations around quasi-equilibrium points as in solids and diffusive motions between different points. The second step was using the above microscopic dynamics to ascertain the nature of excitations in liquids. At the fundamental level, physics of an interacting system is set by its excitations or quasiparticles \cite{landaustat}. In solids, these are phonons. The nature of phonons and their properties in liquids were not clear for a long time since Sommerfeld first brought up this issue in 1913 \cite{somm} (see, e.g., Ref. \cite{zwanzigtau}). A fairly recent combination of theory, experiments and modelling led to understanding the propagation of phonons in liquids with an important property: the phase space available to these phonons is not fixed as in solids but is instead variable \cite{ropp,proctor1,proctor2,chen-review}. This is a non-perturbative effect. In particular, the phonon space in liquids reduces with temperature, consistent with the result from the numerical instantaneous mode approach \cite{keyes}. This reduction has a general implication for liquid thermodynamic properties: specific heat of classical liquids universally decreases with temperature, in agreement with experiments \cite{ropp,proctor1,proctor2}. (In other approaches, the reduction of specific heat was attributed to the singularity of the hard-sphere free energy functional \cite{rosentar} or accounted for by considering the liquid energy as the weighted sum of solid and gas energies, with weights numerically calculated from instantaneous normal modes \cite{instaegami}).

The theory leading to this picture is importantly based on considering the microscopic dynamics of liquid molecules. As discussed in the next section, considering this dynamics is also the key to understanding viscosity minima and calculating their values.

We note in passing that the energy of quantum liquids such as $^4$He is readily understood on the basis of phonons. A quantum nature of this liquid interestingly turns out to be a simplifying circumstance: any weakly perturbed quantum state is a set of elementary quantum excitations. In Bose liquids, excitations can appear and disappear singly (in contrast to Fermi liquids where excitations appear and disappear in pairs). The elementary excitations with small momenta $p$ are the sound waves, the phonons, with the linear dispersion relation $\epsilon=pc$. Hence at temperature close to zero, the elementary excitations are phonons, and the system energy is then the sum of these excitations, resulting in $c_v\propto T^3$ as in solids and in agreement with experiments. Landau attributes this calculation to Migdal in 1940 \cite{landaumigdal}.

\subsection{Viscosity and dynamical crossover}
\label{minimum}

We now discuss the microscopic origin of viscosity minima related to the crossover of particle dynamics.

Viscosity of fluids, $\eta$, varies in a wide range, from about 10$^{-6}$ Pa$\cdot$s for the normal component of liquid He \cite{hevisc} to about 10$^{13}$ Pa$\cdot$s in viscous liquids approaching liquid-glass transition at the glass transition temperature $T_g$. $\eta$ continues to increase below $T_g$ too, however the corresponding relaxation time becomes longer than experimental time. In the low-temperature liquidlike classical regime, $\eta$ has no upper bound as a function of temperature. At temperature approaching zero, $\eta$ is limited by the temperature-independent frequency of particle tunneling.

$\eta$ strongly (exponentially or faster) depends on temperature and pressure. $\eta$ is additionally strongly system-dependent and is governed by the activation energy barrier for molecular rearrangements, $U$. In turn, $U$ is related to inter-molecular interactions and structure. This relationship is in generally complicated, and no universal way to predict $U$ and $\eta$ from first principles exists. Indeed, tractable theoretical models describe the dilute gas limit of fluids where perturbation theory applies, but not dense liquids of interest here \cite{chapman} (in field theories, viscosity can be evaluated in the limit of weak and strong coupling \cite{fieldvis1,fieldvis2}). In view of this and more fundamental problems involved in liquid theory discussed in the previous section, it is quite remarkable that the {\it minimal} value of liquid viscosity turns out to be nearly universal and set by the fundamental physical constants.

\begin{figure}
\begin{center}
{\scalebox{0.37}{\includegraphics{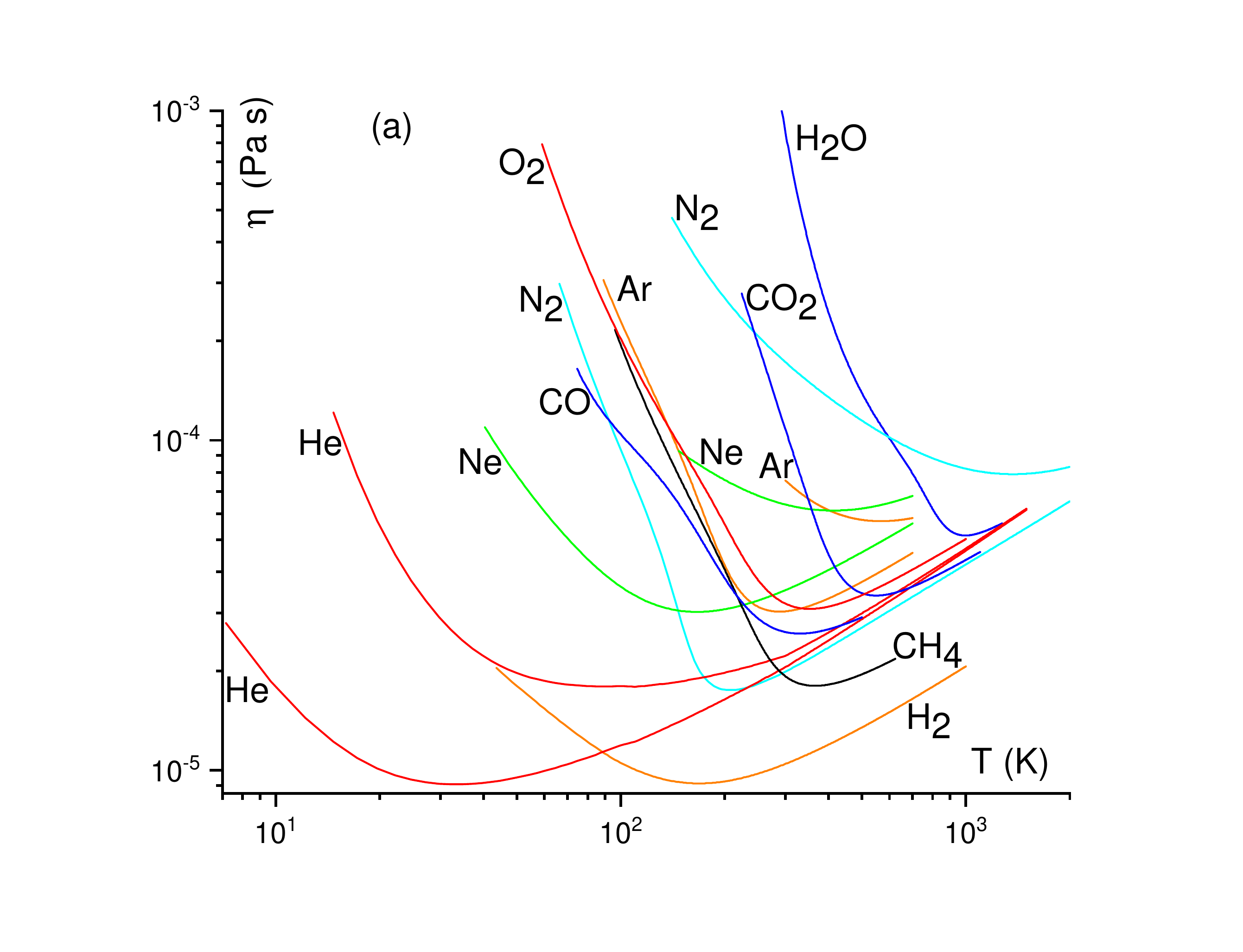}}}
{\scalebox{0.37}{\includegraphics{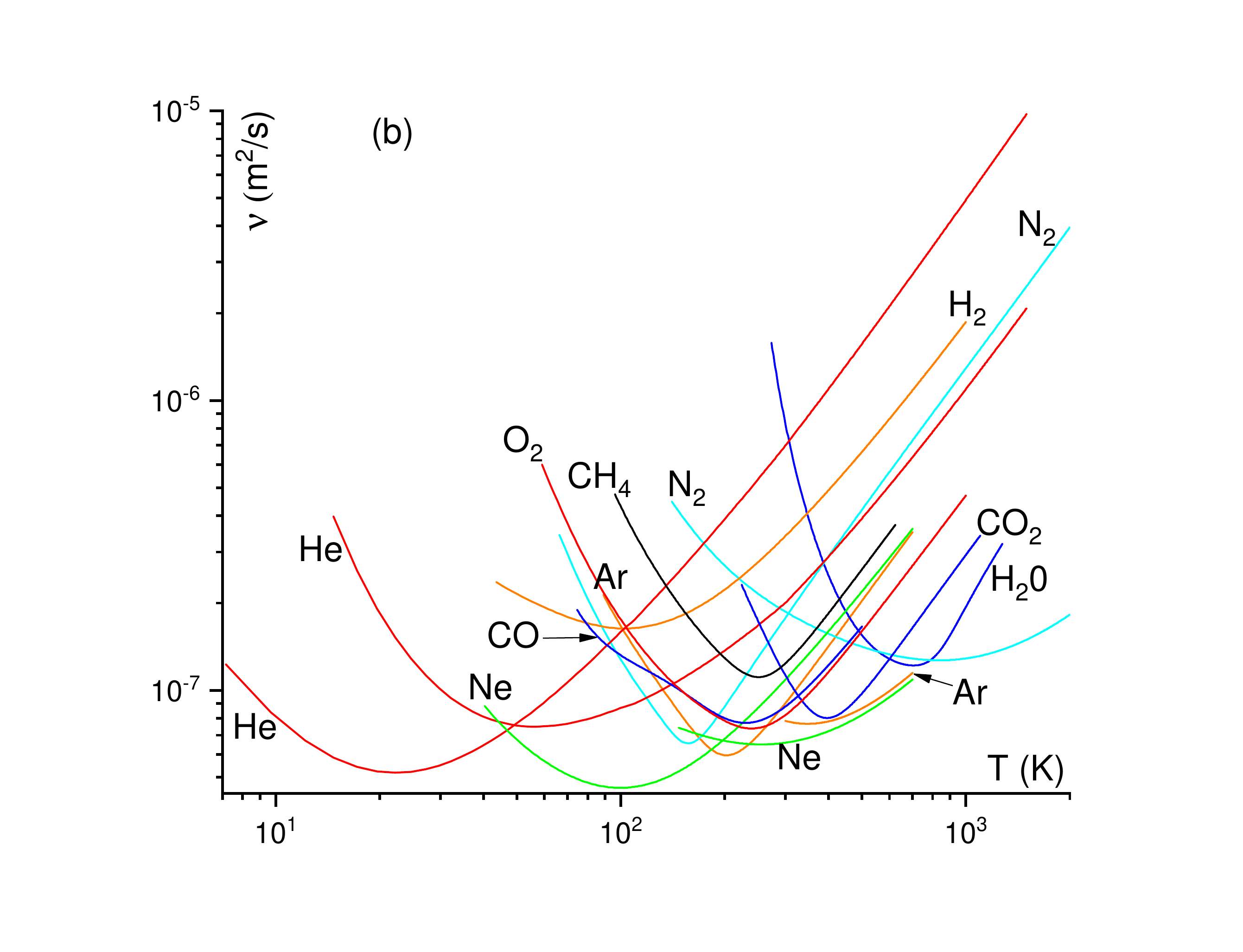}}}
\end{center}
\caption{Experimental dynamic viscosity $\eta$ (a) and experimental kinematic viscosity $\nu$ (b) of noble, molecular and network liquids \cite{nist} showing minima. $\eta$ for H$_2$, H$_2$O and CH$_4$ are shown for pressure $P=50$ MPa, 100 MPa and 20 MPa, respectively. $\eta$ for He, Ne, Ar and N$_2$ are shown at two pressures each: 20 and 100 MPa for He, 50 and 300 MPa for Ne, 20 and 100 MPa for Ar and 10 and 500 MPa for N$_2$. The minimum at higher pressure is above the minimum at lower pressure for each fluid.
}

\label{visc}
\end{figure}

Experimental viscosity $\eta$ and kinematic viscosity $\nu=\frac{\eta}{\rho}$, where $\rho$ is density, are shown in Figure \ref{visc} for several noble (Ar, Ne and He), molecular (H$_2$, N$_2$, CO$_2$, CH$_4$, O$_2$ and CO) and network (H$_2$O) fluids. For some fluids, we show $\eta$ and $\nu$ at two different pressures. The low pressure was chosen to be high enough and above the critical pressure so that viscosity is not affected by near-critical anomalies. The high pressure was chosen to make the considered pressure range as wide as possible and at the same time low enough in order to see the viscosity minima in the temperature range available experimentally.

We now recall the origin of viscosity minima shown in Fig. \ref{visc}. In the liquid-like regime of molecular dynamics at low temperature, $\eta$ decreases with temperature as

\begin{equation}
\eta=\eta_0\exp\left(\frac{U}{T}\right)
\label{v1}
\end{equation}

\noindent where $\eta_0$ is a pre-factor and $U$ can be temperature dependent

In the gas-like regime of molecular dynamics, $\eta$ is

\begin{equation}
\eta=\frac{1}{3}\rho v L
\label{v2}
\end{equation}

\noindent where $\rho$ is density, $v$ is average particle velocity and $L$ is the particle mean free path.

For gases, $L\propto\frac{1}{\rho}$ and $\eta\propto v\propto\sqrt{T}$ \cite{chapman}. Hence $\eta$ increases with temperature without bound, although new effects such as ionization start operating at higher temperature. These can change the system properties including $\eta$.

Consistent with Fig. \ref{visc}, Eqs. (\ref{v1}) and (\ref{v2}) imply that $\eta$ should have a minimum.

Before calculating $\eta$ at the minimum, it is useful to qualify the above terms ``liquid-like'' and ``gas-like'' referring to different regimes of molecular dynamics and elaborate on conditions at which the minima are seen. At low temperature, molecular motion in liquids combines solid-like oscillations around quasi-equilibrium positions and diffusive jumps to new positions. Enabling liquid flow, these jumps are thermally-activated events involving an energy barrier set by inter-molecular interactions. This gives an exponential dependence in Eq. \eqref{v1}. The diffusive jumps are characterised by liquid relaxation time, $\tau$, the average time between the jumps. $\tau$ is related to $\eta$ by the Maxwell relationship $\eta=G\tau$, where $G$ is the high-frequency shear modulus \cite{frenkel}. $\tau$ decreases with temperature in the same way as $\eta$ in Eq. \eqref{v1} and is bound by the elementary vibration period, commonly approximated by the Debye vibration period in the Debye model, $\tau_{\rm D}$. When $\tau$ approaches $\tau_{\rm D}$, the oscillatory component of molecular motion is lost, and molecules start moving in a purely diffusive manner. On further temperature increase (or density decrease), the motion remains purely diffusive, however molecules gain enough energy to move distance $L$ without collisions. In this gas-like regime, the fluid viscosity can be calculated by assuming that a molecule moves in straight lines between collisions, resulting in Eq. \eqref{v2}.

If temperature is increased at pressure below the critical point, the system crosses the boiling line and undergoes the liquid-gas phase transition. As a result, $\eta$ undergoes a sharp change at the transition (we will return to this in Section \ref{elementary}), rather than a smooth minimum as in Fig. \ref{visc}. In order to avoid effects related to the phase transition itself, it is convenient to consider matter above the critical point, the supercritical state. Here, the supercritical Frenkel line (FL) formalises the qualitative change of molecular dynamics from combined oscillatory and diffusive to purely diffusive. Introduced about ten years ago \cite{brazhkin2012,Brazhkin2012a,Brazhkin2013}, the transitions at the FL has been confirmed in several important supercritical fluids using different experimental techniques (see Ref. \cite{flreview} for review). The location of the minima of $\eta$ can depend on the path taken on the phase diagram. As a result, the minimum of $\eta$ may deviate from the FL depending on the path.

\subsection{Viscosity minima}
\label{vminima}

We are now set to calculate viscosity at the minimum, $\eta_{min}$. There are two ways in which this can be done: considering the low-temperature limit of the gas-like viscosity \eqref{v2} or taking the high-temperature limit of the liquid-like viscosity given by the Maxwell relation $\eta=G\tau$. We start with the first approach and consider how $\eta=\rho vL$ changes with temperature decrease (we drop $\frac{1}{3}$ in (\ref{v2}) since the calculation evaluates the order of magnitude of viscosity minimum as discussed in more detail below). $L$ decreases on lowering the temperature and is bound by the UV cutoff in condensed matter systems: inter-particle separation $a$. From this point on, $L$ has no room to decrease further. Instead, the system enters the liquid-like regime where $\eta$ starts increasing on further temperature decrease according to (\ref{v1}) because the diffusive molecular motion crosses over to thermally-activated as discussed earlier. Therefore, $\eta_{min}$ approximately corresponds to $L\approx a$. When $L$ becomes comparable to $a$, $v$ in Eq. \eqref{v2} can be evaluated as $v=\frac{a}{\tau_{\rm D}}$ because the time for a molecule to move distance $a$ in this diffusive regime is given by the characteristic time scale set by $\tau_{\rm D}$. Setting $L=a$, $v=\frac{a}{\tau_{\rm D}}=\frac{1}{2\pi}\omega_{\rm D}a$ and $\rho\approx\frac{m}{a^3}$, where $\omega_{\rm D}$ is Debye frequency and $m$ is molecule mass, gives:

\begin{equation}
\eta_{min}=\frac{1}{2\pi}\frac{m\omega_{\rm D}}{a}
\label{v3}
\end{equation}

We note that (\ref{v2}) applies in the regime where $L$ is larger than $a$, hence the evaluation of viscosity minimum is an order-of-magnitude estimation. This is consistent with other approximations made later. In this regard, we observe that theoretical models can only describe viscosity in a dilute gas limit where perturbation theory applies \cite{chapman}, but not in the regime where $L\approx a$ and where the energy of inter-molecular interaction is comparable to the kinetic energy. In view of theoretical issues as well as many orders of magnitude by which $\eta$ can vary, the evaluation of its minimum is meaningful and informative. An order-of-magnitude evaluation is probably unavoidable if a complicated property such as viscosity is to be expressed in terms of fundamental constants only.

$\eta_{min}$ in (\ref{v3}) matches the result obtained by approaching the viscosity minimum from low temperature in the liquid-like regime and considering the Maxwell relationship $\eta=G\tau$. In the liquid-like regime, $\eta$ and $\tau$ decrease with temperature according to (\ref{v1}), but this decrease is bound from below because $\tau$ starts approaching the shortest time scale in the system set by the Debye vibration period, $\tau_{\rm D}$. From this point on, $\tau$ has no room to decrease further, and the system enters the gas-like regime where $\eta$ starts increasing with temperature according to (\ref{v2}). This corresponds to the crossover between the thermally-activated liquid-like and diffusive gas-like motion of molecules discussed earlier. Therefore, the minimum of $\eta$ can be evaluated by setting $\tau\approx\tau_{\rm D}$. In the liquid-like regime, $G$ can be estimated as $G=\rho c^2$, where $c\approx\frac{a}{\tau_{\rm D}}$ is the speed of sound. Then, $\eta_{min}=G\tau_{\rm D}=\rho\frac{a^2}{\tau_{\rm D}}=\frac{1}{2\pi}\frac{m\omega_{\rm D}}{a}$ as in Eq. (\ref{v3}), where $\rho=\frac{m}{a^3}$ is used as before.

We can check how well Eq. (\ref{v3}) evaluates the minima of $\eta$ in Figure \ref{visc}. Taking characteristic values $a=$3-6 \AA, $\frac{\omega_{\rm D}}{2\pi}$ on the order of 1 THz and atomic weights 2-40 for liquids in Fig. \ref{visc}, we find $\eta_{min}$ in the range $10^{-5}-10^{-4}$ Pa$\cdot$s. This is consistent with Fig. 1a. We observe that high pressure reduces $a$ and increases $\omega_{\rm D}$. As a result, Eq. (\ref{v3}) predicts that $\eta_{min}$ increases with pressure, in agreement with the experimental behavior in Fig. \ref{visc}a.

The viscosity minima of strongly-bonded liquids such as liquids metals were not measured due to their high critical points. Nevertheless, high-temperature $\eta$ is close to $10^{-3}$ Pa$\cdot$s for Fe (2000 K), Zn (1100 K), Bi (1050 K) \cite{metals}, Hg (573 K) and Pb (1173 K) and is expected to be close to $\eta$ at the minima. This is larger than $\eta_{min}$ in Fig. \ref{visc} and is consistent with Eq. (\ref{v3}) predicting that $\eta_{min}$ decreases with $a$ ($a$ is smaller in metallic systems as compared to noble and molecular ones in Fig. \ref{visc}a) and increases with molecular mass ($m\omega_{\rm D}\propto\sqrt{m}$).

It is convenient to use the kinematic viscosity $\nu$.
$\nu$ describes momentum diffusivity, analogous to thermal diffusivity involved in heat transfer discussed in Chapter \ref{thermal} and gives the diffusion constant in the gas-like regime of molecular dynamics \cite{frenkel}. Another benefit of considering $\nu$ is that it makes the link to the high-energy result discussed in Section \ref{qgp}, where $\eta$ is divided by the volume density of entropy. Using $\nu=\frac{\eta}{\rho}=vL$, $v=\frac{1}{2\pi}a\omega_{\rm D}$ and $L=a$ as before gives the minimal value of $\nu$, $\nu_{min}$, as

\begin{equation}
\nu_{min}=\frac{1}{2\pi}\omega_{\rm D}a^2
\label{nu}
\end{equation}

We now come to an important part of this discussion where we invoke fundamental physical constants \cite{sciadv1}. We recall that the properties defining the UV cutoff in condensed matter can be expressed in terms of these constants. Two important properties are Bohr radius, $a_{\rm B}$, setting the characteristic scale of inter-particle separation in condensed matter phases on the order of Angstrom:

\begin{equation}
a_{\rm B}=\frac{4\pi\epsilon_0\hbar^2}{m_e e^2}
\label{bohr}
\end{equation}

\noindent and the Rydberg energy, $E_{\rm R}=\frac{e^2}{8\pi\epsilon_0a_{\rm B}}$ \cite{ashcroft}, setting the characteristic scale of cohesive energy in condensed matter phases on the order of several eV:

\begin{equation}
E_{\rm R}=\frac{m_ee^4}{32\pi^2\epsilon_0^2\hbar^2}
\label{rydberg}
\end{equation}

\noindent where $e$ and $m_e$ are electron charge and mass.

Lets now recall the known ratio between the cohesive energy $E$ and the characteristic phonon energy, $\hbar\omega_{\rm D}$: $\frac{\hbar\omega_{\rm D}}{E}$. This ratio can be derived by approximating $\hbar\omega_{\rm D}$ as $\hbar\left(\frac{E}{ma^2}\right)^{\frac{1}{2}}$, taking the ratio $\frac{\hbar\omega_{\rm D}}{E}$ and using $a=a_{\rm B}$ from (\ref{bohr}) and $E=E_{\rm R}$ from (\ref{rydberg}). This gives, up to a factor close to 1:

\begin{equation}
\frac{\hbar\omega_{\rm D}}{E}=\left(\frac{m_e}{m}\right)^{\frac{1}{2}}
\label{ratio}
\end{equation}

The same ratio \eqref{ratio} follows by combining two known relations in metallic systems: $\frac{\hbar\omega_{\rm D}}{E}\approx\frac{c}{v_F}$, where $v_F$ is the Fermi velocity, and $\frac{c}{v_F}\approx\left(\frac{m_e}{m}\right)^{\frac{1}{2}}$, providing an order-of-magnitude estimation $\frac{\hbar\omega_{\rm D}}{E}$ in other systems too \cite{ashcroft}.

Combining (\ref{nu}) and (\ref{ratio}) gives

\begin{equation}
\nu_{min}=\frac{1}{2\pi}\frac{E a^2}{\hbar}\left(\frac{m_e}{m}\right)^{\frac{1}{2}}
\label{nu01}
\end{equation}

As mentioned earlier, $a$ and $E$ in (\ref{nu01}) are set by their characteristic values $a_{\rm B}$ and $E_{\rm R}$. Using $a=a_{\rm B}$ from (\ref{bohr}) and $E=E_{\rm R}$ from (\ref{rydberg}) in (\ref{nu01}) gives a simple and good-looking result for $\nu_{min}$:

\begin{equation}
\nu_{min}=\frac{1}{4\pi}\frac{\hbar}{\sqrt{m_em}}
\label{nu1}
\end{equation}

Eq. \eqref{nu1} can be obtained without explicitly using $a_{\rm B}$ and $E_{\rm R}$ in (\ref{nu01}). The cohesive energy, or the characteristic energy of electromagnetic interaction, is

\begin{equation}
E=\frac{\hbar^2}{2m_ea^2}
\label{direct}
\end{equation}

Using (\ref{direct}) in (\ref{nu01}) gives (\ref{nu1}).

Another way to derive (\ref{nu1}) is to consider the ``characteristic'' viscosity $\eta^*$ \cite{reduced}:

\begin{equation}
\eta^*=\frac{(E m)^\frac{1}{2}}{a^2}
\label{etastar}
\end{equation}

$\eta^*$ is used to describe viscosity scaling on the phase diagram: the ratio between viscosity and $\eta^*$ is the same for systems described by the same interaction potential in equivalent points of the phase diagram. For systems described by the Lennard-Jones potential, the experimental and calculated viscosity near the triple point and close to the melting line is about 3 times larger than $\eta^*$ \cite{reduced,vadim}. Near the critical point, $\eta^*$ is about 4 times larger than viscosity and is expected to give the right order of magnitude of viscosity at the minimum at moderate pressure. The kinematic viscosity corresponding to (\ref{etastar}) is

\begin{equation}
\frac{\eta^*}{\rho}=\frac{E^\frac{1}{2}a}{m^{\frac{1}{2}}}
\label{etastar1}
\end{equation}

Using $a=a_{\rm B}$ from (\ref{bohr}) and $E=E_{\rm R}$ from (\ref{rydberg}) in (\ref{etastar1}) gives the same result as (\ref{nu1}) up to a constant factor on the order of unity. As before, we can also use (\ref{direct}) in (\ref{etastar1}) to get the same result.

Minimal viscosity in Eq. \eqref{nu1} corresponds to maximal fluidity in the system.

We observe that $\nu_{min}$ in \eqref{nu1} contains $\hbar$ and electron and molecule masses only. Lets consider the implications of this in more detail.

The first observation is that viscosity is commonly considered as a classical property because most liquids exist at high temperature and are classical. This is related to melting temperature exceeding the Debye temperature in most systems. Yet the minimal viscosity is a quantum property as follows from Eq. \eqref{nu1}. This is because viscosity it is governed by molecular interactions, and these are ultimately set by quantum effects. Brazhkin has expanded on this point in relation to viscosity and other properties of condensed matter \cite{brazhkinreview}.

Second, $\nu_{min}$ interestingly does not depend on electron charge $e$, contrary to what one might expect considering that viscosity is set by the inter-particle forces which are electromagnetic in origin. Although $e$ enters Eqs. \eqref{bohr}, \eqref{rydberg} and \eqref{direct} for the Bohr radius, Rydberg and cohesive energy, it cancels out in Eq. \eqref{nu1} for $\nu_{min}$. We will return to this point later in section \ref{qgp}.

Third, there are two masses in Eq. \eqref{nu1}, $m$ and $m_e$. $m$ characterises the molecules involved in viscous flow. $m_e$ characterises electrons setting the inter-molecular interactions. $m$ in (\ref{nu1}) is $m=Am_p$, where $A$ is the atomic weight and $m_p$ is the proton mass. The inverse square root dependence $\nu_{min}\propto\frac{1}{\sqrt{A}}$ interestingly implies that $\nu_{min}$ is not too sensitive to the liquid type.

Setting $m=m_p$ ($A=1$) for H in (\ref{nu1}) (similarly to (\ref{bohr}) and (\ref{rydberg}) derived for the H atom) gives the fundamental kinematic viscosity $\nu_f$ in terms of $\hbar$, $m_e$ and $m_p$ as

\begin{equation}
\nu_f=\frac{1}{4\pi}\frac{\hbar}{\sqrt{m_em_p}}\approx 10^{-7} \frac{\rm{m}^2}{\rm{s}}
\label{nuf}
\end{equation}

Eq. \eqref{nuf} is consistent with the experimental results in Figure \ref{visc}b. This shows how fundamental constants set the characteristic scale of physical properties. This includes complicated properties such as viscosity which was not thought to be amenable to an analytical treatment. We will revisit this point in Section \ref{weinbergfun}.

We note that a relationship between fundamental constants and simpler properties such as elastic moduli discussed in Chapter \ref{elastic} was known and is not unexpected. For more complicated properties such as viscosity discussed here, thermal diffusivity and speed of sound discussed in Chapters \ref{thermal} and \ref{sound}, it remained unclear till fairly recently whether their characteristic values can be directly related to fundamental constants. One of the aims of this review is to show how this can be done.

\begin{figure}
\begin{center}
{\scalebox{0.25}{\includegraphics{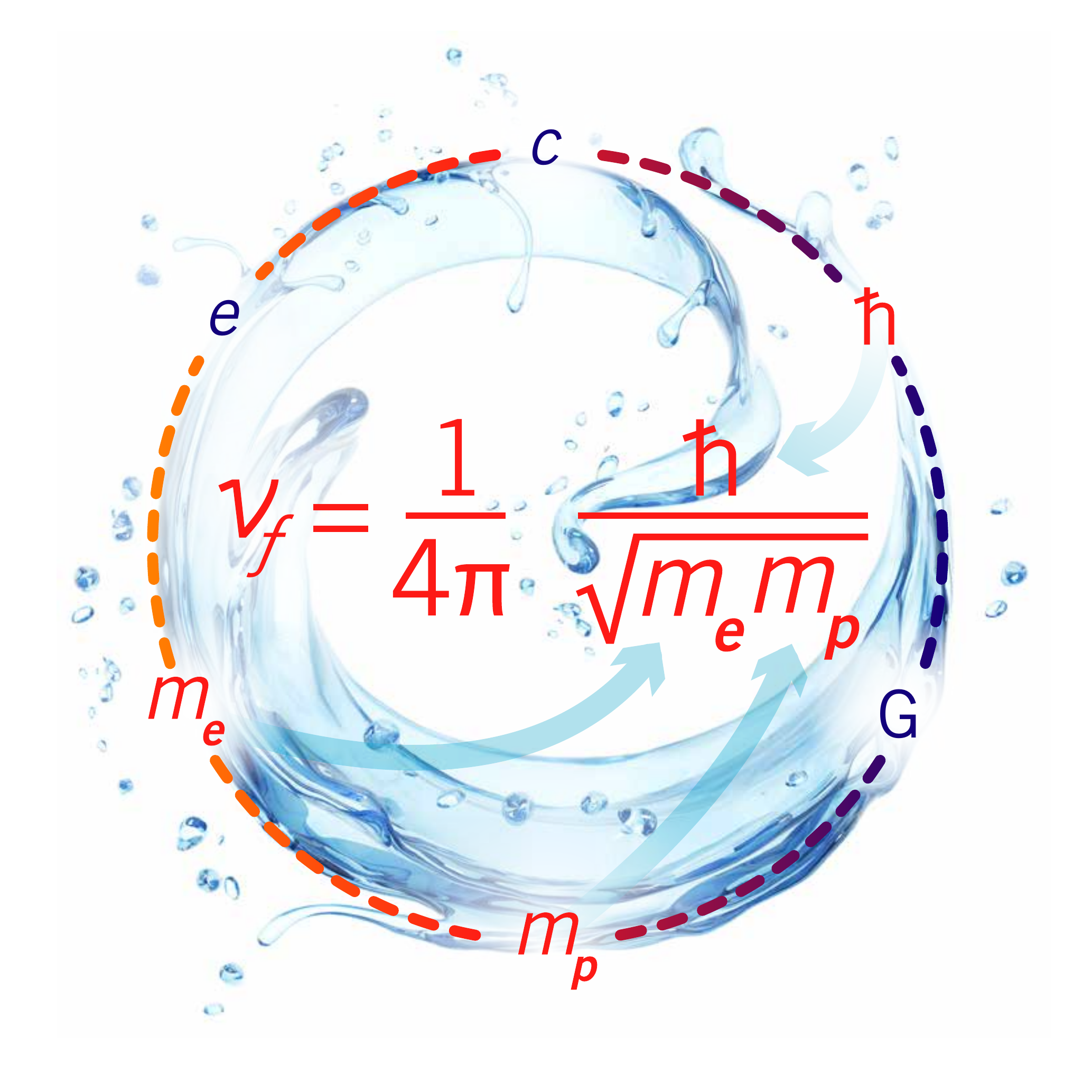}}}
\end{center}
\caption{Fundamental kinematic viscosity depends on three values only: $\hbar$, $m_e$ and $m_p$.
}
\label{fundamental}
\end{figure}

$\nu_f$ depends on three parameters: $\hbar$, $m_e$ and $m_p$, as illustrated in Figure \ref{fundamental}. $\hbar$ and $m_e$ are fundamental constants. Although $m_p$ depends on other Standard Model parameters, the dimensionless number $\frac{m_e}{m_p}$ is attributed a fundamental importance \cite{barrow}, as discussed in section \ref{life} in more detail.

The derivation of the viscosity minimum \eqref{nu1} and fundamental viscosity \eqref{nuf} involves more than a dimensional analysis. First, the dimensionless analysis is not unique in the absence of a physical model. As mentioned earlier, it is not apriori clear that $\nu_{min}$ should involve $\hbar$ and be a quantum property, especially so in view that most liquids are considered classical. Second, a purely dimensional analysis can give a quantity with right dimensions but wrong value. Indeed, multiplying the right hand side of Eq. \eqref{nu1} by $f(\frac{m}{m_e})$, where $f$ is an arbitrary function, gives the result consistent with the dimensional analysis but produces any desired value of $\nu_{min}$ with a suitable choice of $f$. Third, we have used a specific physical model to derive $\nu_{min}$. We started with attributing the viscosity minimum to the crossover of microscopic particle dynamics, from combined oscillatory and diffusive to purely diffusive. This consideration led us to a particular regime of particle dynamics where the particle speed is set by the interatomic separation and elementary vibration period. Dimensional analysis alone does not have anything to say about why would this regime correspond to the {\it minimum} of viscosity. We next evaluated the minimal value of $\eta_{min}$ using two approaches involving the Maxwell relation and the gas kinetic theory. Each of these approaches is based on a specific physical mechanism. We then related $\nu_{min}$ to the length and energy scales using the ratio $\frac{\hbar\omega_{\rm D}}{E}=\left(\frac{m_e}{m}\right)^{\frac{1}{2}}$ in Eq. \eqref{ratio}. The dimensionality analysis does not predict this ratio and is consistent with $\frac{\hbar\omega_{\rm D}}{E}$ taking any dimensionless number. We finally expressed the length and energy scales in terms of fundamental constants. Most of these steps involved in the derivation of Eq. \eqref{nu1} are physically guided and incorporate a lot more information that would be available from purely dimensional considerations.

In Table \ref{table}, we compare $\nu_{min}$ calculated according to the theoretical prediction (\ref{nu1}) to the experimental $\nu_{min}$ \cite{nist} for all systems shown in Fig. \ref{visc}. The ratio between experimental and predicted $\nu_{min}$ is in the range of 0.5-3. As expected, experimental $\nu_{min}$ for the lightest liquid in Table \ref{table}, H$_2$, is close to the theoretical fundamental viscosity (\ref{nuf}). In view of approximations made, we observe that Eq. (\ref{nu1}) predicts $\nu_{min}$ well.

\begin{table}[ht]
\begin{tabular}{ l l l}
\hline
                   & $\nu_{min}$ (calc.)       & $\nu_{min}$ (exp.)\\
                   & $\times$10$^8$ m$^2$/s & $\times$10$^8$ m$^2$/s\\
\hline
Ar (20 MPa)        & 3.4        & 5.9 \\
Ar (100 MPa)       & 3.4        & 7.7 \\
Ne (50 MPa)        & 4.8        & 4.6 \\
Ne (300 MPa)       & 4.8        & 6.5 \\
He (20 MPa)        & 10.7       & 5.2 \\
He (100 MPa)       & 10.7       & 7.5 \\
N$_2$ (10 MPa)     & 4.1        & 6.5 \\
N$_2$ (500 MPa)    & 4.1        & 12.7 \\
H$_2$ (50 MPa)     & 15.2       & 16.3 \\
O$_2$ (30 MPa)     & 3.8        & 7.4 \\
H$_2$O (100 MPa)   & 5.1        & 12.1 \\
CO$_2$ (30 MPa)    & 3.2        & 8.0 \\
CH$_4$ (20 MPa)    & 5.4        & 11.0 \\
CO (30 MPa)        & 4.1        & 7.7 \\
\hline
\end{tabular}
\caption{Calculated and experimental $\nu_{min}$.}
\label{table}
\end{table}

Table \ref{table} shows that $\nu_{min}$ increases with pressure in Table \ref{table}, similarly to $\eta_{min}$ in Fig. \ref{visc}. However, pressure dependence is not accounted in $\nu_{min}$ in (\ref{nu1}) since (\ref{nu1}) is derived using Eqs. (\ref{bohr})-(\ref{nu01}) which do not account for the pressure dependence of $\omega_{\rm D}$ and $E$.

We add several other remarks regarding the comparison in Table \ref{table}. First, the important term in Eq. (\ref{nu1}) is the combination of fundamental constants ${\hbar}$, $m_e$ and $m_p$ which set the characteristic scale of the minimal kinematic viscosity, whereas the numerical factor may be affected by the approximations used and mentioned earlier. Second, Eqs. (\ref{bohr})-(\ref{ratio}) assume valence electrons directly involved in chemical bonding and hence strongly-bonded systems, including covalent, ionic and metallic liquids. Their viscosity in the supercritical state is generally unavailable due to high critical points. The available experimental data in Fig. \ref{visc} and Table \ref{table} includes weakly-bonded systems such as noble, molecular and hydrogen-bonded fluids. Although bonding in these systems is also electromagnetic in origin, weaker dipole and van der Waals interactions corresponds to smaller $E$ and, consequently, smaller $\eta$ as compared to strongly-bonded ones, with the viscosity of hydrogen-bonded fluids lying in between \cite{vadim1}. However, $\nu_{min}$ in (\ref{nu01}) contains the factor $Ea^2$. $E^{\frac{1}{2}}$ is 3-10 times smaller and $a$ is 2-4 times larger in weakly-bonded as compared to strongly-bonded systems \cite{vadim1}. Hence the dependence of $\nu_{min}$ on bonding type is weak. As a result, the order-of-magnitude evaluation (\ref{nu1}) is unaffected, as Table 1 shows.

More recently, the experimental viscosity of metallic liquids was discussed at high temperature in order to find limiting high-temperature value of Eq. \eqref{v1}, $\eta_0$, and compare it to the bound \eqref{nu1} \cite{nussinov,nussinov1,nussinov2}. Although the bound \eqref{nu1} is related to the true minimum of viscosity and can be several times larger than $\frac{\eta_0}{\rho}$ due to the crossover between liquidlike and gaslike dynamics, the closeness between the predicted bound \eqref{nu1} and experimental $\frac{\eta_0}{\rho}$ was noted.

\subsection{Elementary viscosity, diffusion constant and uncertainty principle}
\label{elementary}

Corresponding to atomic H, Eq. (\ref{nuf}) gives the maximal value of the minimal kinematic viscosity. It is interesting to find a viscosity-related quantity which has an absolute minimum. This can be done by introducing the ``elementary'' viscosity $\iota$ (``iota'') defined as the product of $\eta_{min}$ and elementary volume $a^3$: $\iota=\eta_{min} a^3$ or, equivalently, as $\iota=\nu_{min}m$. Using (\ref{nu1}), $\iota$ is

\begin{equation}
\iota=\frac{\hbar}{4\pi}\left({\frac{m}{m_e}}\right)^{\frac{1}{2}}
\label{iota1}
\end{equation}

Eq. \ref{iota1} has the absolute lower bound, $\iota_{min}$, for $m=m_p$ in H:

\begin{equation}
\iota_{min}=\frac{\hbar}{4\pi}\left({\frac{m_p}{m_e}}\right)^{\frac{1}{2}}
\label{iota2}
\end{equation}

\noindent which is on the order of $\hbar$ ($\iota_{min}\approx 3.4\hbar$) and interestingly involves the proton-to-electron mass ratio, one of few dimensionless combinations of fundamental constants of general importance \cite{barrow,uzanreview}.

\begin{figure}
\begin{center}
{\scalebox{0.35}{\includegraphics{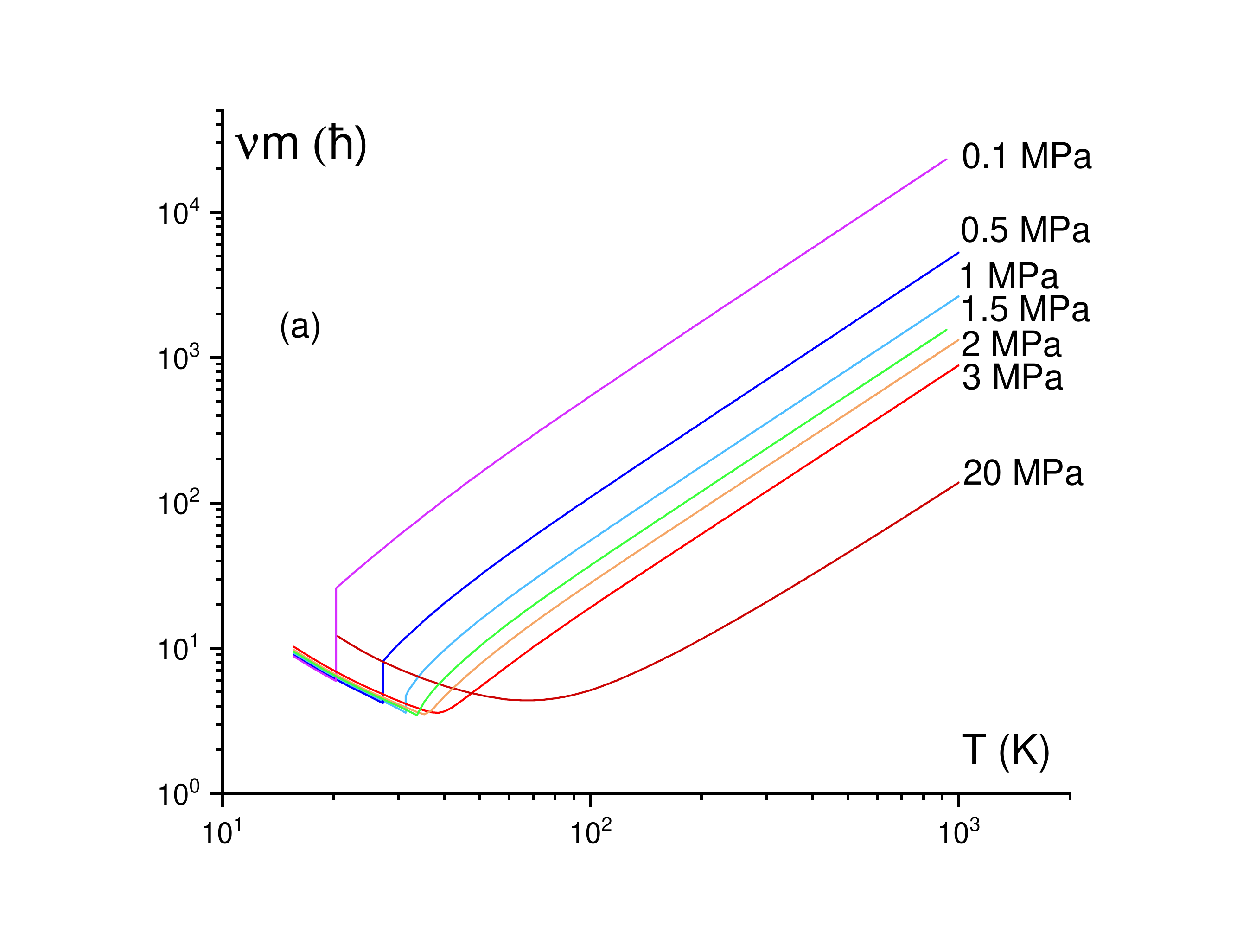}}}
{\scalebox{0.35}{\includegraphics{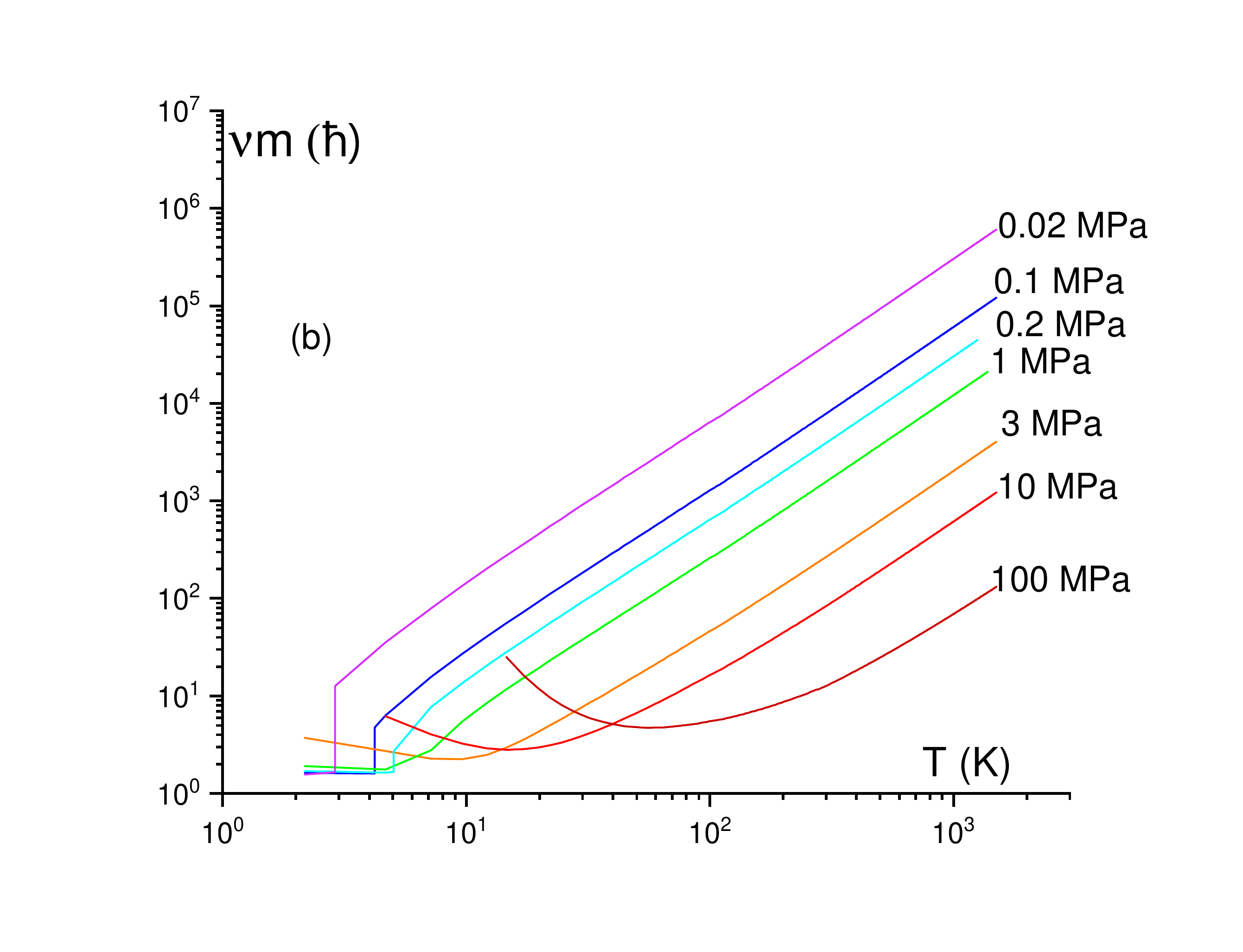}}}
{\scalebox{0.35}{\includegraphics{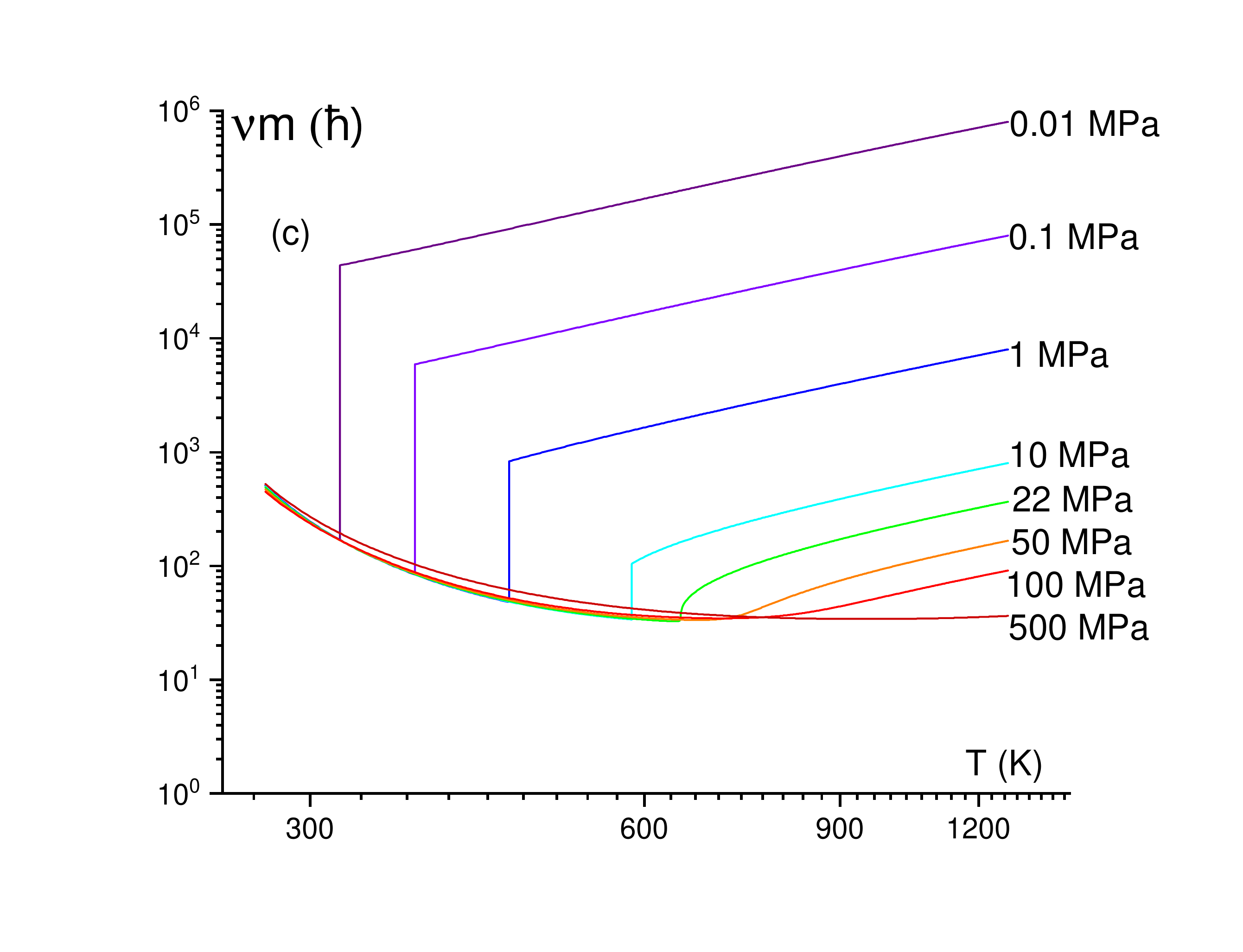}}}
\end{center}
\caption{$\nu m$ calculated from experimental kinematic viscosity \cite{nist} for H$_2$ (a), He (b) and H$_2$O (c) below and above the critical pressure $P_c$. $P_c$=1.3 MPa for H$_2$, 0.23 MPa for He and 22 MPa for H$_2$O. The smallest value of $\nu m$, $\iota$, is consistent the lower bound (\ref{iota2}). From Ref. \cite{sciadv1}. 
}
\label{3}
\end{figure}

In Fig. \ref{3}a-b, we show the product $\nu m$ in the units of $\hbar$ for two lightest liquids, H$_2$ and He, for which the minimum of $\nu m$, $\nu_{min}m=\iota$, should be close to the lower bound (\ref{iota2}). $\nu m$ is calculated using the experimental viscosity and density data \cite{nist} and shown above and below the critical pressure $P_c$. For He, the temperature range is above the superfluid transition (we do not consider superfluidity here).

We observe that the liquid-gas phase transition results in sharp changes of viscosity below $P_c$. For H$_2$, the minimum of $\nu m$ is kinked as a result and, starting from the lower pressure, decreases with pressure up to $P_c$. This is followed by the minimum becoming smooth and increasing above $P_c$. The smooth minimum just above the critical point (where the derivation of $\eta_{min}$ and $\nu_{min}$, assuming a smooth variation of viscosity, applies) is very close to the minimum at $P_c$. For He, the minimum similarly increases with pressure above $P_c$ and weakly varies below $P_c$.

The smallest value of $\nu m$, $\iota=\nu_{min}m$, in Fig. \ref{3}a-b is in the range (1.5-3.5)$\hbar$ for He and H$_2$. This is consistent with the estimation of the lower bound of $\iota$, $\iota_{min}$ in (\ref{iota2}). Given that $\nu m$ varies 4-6 orders of magnitude in Fig. \ref{3}, the agreement with Eq. (\ref{iota2}) is notable.

We also show $\nu m$ for common H$_2$O in Fig. \ref{3}c as a useful reference and include the triple and critical point in the pressure range. The behavior of $\nu m$ is similar to that of H$_2$, with $\iota$ of about 30$\hbar$. Similarly to H$_2$ and He, the smooth minimum just above $P_c$ is very close to that at $P_c$. This implies that the viscosity minimum applies to both supercritical fluids and subcritical liquids. We will return to this point in the next Section.

$\iota$ is a convenient property to discuss the uncertainty principle and its implication for the lower bound. As discussed in the previous section, the minimum of $\nu$ can be evaluated as $\nu_{min}=va$, corresponding to $\iota=mva=pa$, where $p$ is particle momentum. According to the uncertainty principle applied to a particle localised in the region set by $a$, $\iota\ge\hbar$. This is consistent with the bound $\iota_{min}$ in (\ref{iota2}), although a more general Eq. (\ref{iota1}) gives a stronger bound which increases for heavier molecules.

An important difference of the lower bound \eqref{nu1} or \eqref{iota1} and bounds based on the uncertainty relation \cite{behnia,zaanen,maldacena,hartnoll,hartnoll1,spin} or other mechanisms \cite{grozdanov} in earlier discussions  is that \eqref{nu1} and \eqref{iota1} correspond to a true minimum as seen in Fig. \ref{visc} (in a sense that the function has an extremum), whereas the uncertainty relation compares a product ($px$ or $Et$) to $\hbar$ but the product does not necessarily correspond to a minimum of a function and can apply to a monotonic function. We will return to this point in Section \ref{qgp}.

The uncertainty relation can also used to evaluate the diffusion constant $D$ and its lower bound in the gaslike regime of particle dynamics (the upper bound on diffusion constant related to relativistic effects was also discussed \cite{hartupper}). In the gaslike and liquidlike regimes of particle dynamics (see Section \ref{minimum}), $D=\nu$ and $D\propto\frac{1}{\nu}$, respectively \cite{frenkel}. This implies that, differently from $\eta$ and $\nu$, $D$ does not have a minimum and monotonically increases with temperature, albeit with a crossover at the Frenkel line marking the transition from gaslike to liquidlike dynamics as discussed in Section \ref{minimum}. However, the lower bound in the gaslike regime can be found by using the same approach we used for viscosity minimum earlier and by equating the particle mean to $a$: $D_{min}=\nu_{min}=va$. Combining this with the uncertainty relation $pa\ge\hbar$, we find the lower bound of $D$ as

\begin{equation}
D_{min}\ge\frac{\hbar}{m}
\label{dminim}
\end{equation}

Eq. \eqref{dminim} is found to be consistent with diffusion experiments in Fermi gases \cite{Dzwerlein}.

\subsection{The Purcell question: why do all viscosities stop at the same place?}

In 1977, Purcell noted that there is almost no liquid with viscosity much lower than that of water and observed (original italics preserved) \cite{purcell}:

\begin{quotation}
``The viscosities have a big range {\it but they stop at the same place}. I don't understand that.''
\end{quotation}

In the first footnote of that paper, Purcell says that Weisskopf has explained this to him. We did not find published Weisskopf's explanation, however the same year Weisskopf published the paper ``About liquids'' \cite{weisskopf}. That paper starts with a story often recited by conference speakers: imagine a group of isolated theoretical physicists trying to deduce the states of matter using quantum mechanics only. They are able to predict the existence of gases and solids, but not liquids.

Earlier discussion in this Chapter helps answer the Purcell question. The answer has two parts. First, viscosities ``stop'' because they have minima. Second, the minima are fairly fixed by fundamental physical constants: these constants help keep $\nu_{min}$ in Eq. \eqref{nu1} from moving up or down too much \cite{pt2021}. $\nu_{min}$ are not universal due to $\nu_{min}\propto\frac{1}{\sqrt{m}}$ mass dependence, although this does change $\nu_{min}$ too much for most liquids. This includes liquids listed in Table \ref{table}.

For different fluids such as those in Fig. \ref{visc} and Table \ref{table}, Eq. (\ref{nu1}) predicts $\nu_{min}$ in the range (0.3-1.5)$\cdot 10^{-7}\frac{{\rm m}^2}{\rm s}$. This is somewhat lower, but not far, from $\nu$ in water at room conditions. Water at ambient conditions happens to be runny enough and close to the minimum. This is what Purcell noted: viscosities of most liquids do not go much lower than in water.

An interesting implication of this discussion is related to our everyday experience in which we deal with water and water-based substances. We have earlier seen that water viscosity is not far from what Eq. \eqref{nuf} predicts. This prompts an interesting thought: our daily experience is set by three fundamental constants in Eq. \eqref{nuf}. We will find similar examples later on in this review.

In Section \ref{vminima}, the lower viscosity bound $\eta_{min}$ or $\nu_{min}$ was related to a smooth viscosity minimum such as that shown in Figure \ref{visc}. The smoothness was due to the crossover in the supercritical state where no liquid-gas phase transition intervenes. If we are below the critical point, $\nu$ still has a minimum, albeit with a jump as is seen in Figure \ref{3}. This Figure also shows that the smallest value of all minima involving jumps below the critical pressure nearly coincides with the low-lying smooth minimum above the critical point. Therefore, the lower viscosity bound applies to both the subcritical and supercritical liquids. This is relevant to the Purcell question: although he did not specify which liquids he examined, he was probably referring mostly to subcritical liquids.

\subsection{Fundamental constants, quantumness and life}
\label{life}

We recall the fundamental physical constants appearing in Eq. \eqref{nu1} and Eq. \eqref{nuf} including $\hbar$ and $m_e$. These and other constants form dimensionless fundamental constants which do not depend on the choice of units and which play a special role in physics \cite{barrow}. Two important numbers are the fine structure constant $\alpha=\frac{e^2}{\hbar c}$ and the electron-to-proton mass ratio, $\frac{m_e}{m_p}$. The finely-tuned values of $\alpha$ and $\frac{m_e}{m_p}$, and the balance between them, governs nuclear reactions and nuclear synthesis in stars, leading to the creation of the essential biochemical elements, including carbon, and molecular structures essential to life. This balance provides a narrow ``habitable zone'' in the ($\alpha$,$\frac{m_e}{m_p}$) space where stars and planets can form and life-supporting molecular structures can emerge \cite{barrow}. For this reason, Barrow calls these constants ``bio-friendly'' and Adams refers to our Universe as ``biophilic''. Adams gives a detailed review of fine-tuning of fundamental constants in our Universe and possible other Universes \cite{adamsreview}. Focusing primarily on astronomy, cosmology and particle physics, their review discusses variations of fundamental constants which can support life.

On the basis of Eq. \eqref{nu1} or Eq. \eqref{nuf}, we can add another observation. The currently observed fundamental constants are friendly to life at a higher level too: biological processes, including those in cells and inter-cellular processes rely heavily on water. Lets consider what would happen if fundamental constants were to take different values. According to Eq. \eqref{nu1}, the minimum of $\nu$ and hence the minimum of $\eta$ will change accordingly. To be more specific, let's write the linearised Navier-Stokes equation as

\begin{equation}
\rho\frac{\partial {\bf v}}{\partial t}=-\grad p+\eta\nabla^2{\bf v}
\label{navier}
\end{equation}

\noindent where ${\bf v}$ is the fluid velocity which is assumed to be small and $p$ is pressure.

For time-dependent flow, the solution of Eq. \eqref{navier} depends on kinematic viscosity $\frac{\eta}{\rho}$. For simplicity, we consider steady flow where the flow velocity depends on $\eta$. Using $\eta_{min}=\nu_{min}\rho$, $\rho\propto\frac{m}{a_{\rm B}^3}$, $m\propto m_p$ and Eqs. \eqref{bohr} and \eqref{nu1}, we find

\begin{equation}
\eta_{min}\propto\frac{e^6}{\hbar^5}\sqrt{m_pm_e^5}
\label{etamin}
\end{equation}

Let's consider diffusive processes in and between cells. These processes correspond to the low-temperature liquidlike dynamics involving combined oscillatory and diffusive particle motion (see Section \ref{minimum}). In this regime, the Stokes-Einstein equation relates $\eta$ and diffusion constant $D$ as \cite{frenkel}:

\begin{equation}
D=\frac{T}{6\pi r\eta}
\end{equation}

\noindent where $r$ is the radius of moving particle.

The minimal viscosity in Eq. \eqref{etamin} then gives the largest $D$ attainable in the system and limits diffusion from above.

Lets consider what happens if we dial $\hbar$ and set it smaller than the current value. $\eta_{min}$ in Eq. \eqref{etamin} is quite sensitive to $\hbar$ and increases if $\hbar$ is smaller. Raising the viscosity minimum implies that viscosity of {\it all} liquids increases, at {\it all} conditions of pressure and temperature. Larger viscosity means that water now flows slower, dramatically affecting life processes such as blood flow, vital flow processes in cells, inter-cellular processes and so on. At the same time, diffusion strongly decreases, implying slowing down of all diffusive processes of essential substances and molecular structures in and across cells. This affects, for example, protein mobility, active transport involving protein motors and cytoskeletal filaments, molecular transport, cytoplasmic mixing, mobility of cytoplasmic constituents and sets the limits at which molecular interactions and biological reactions can occur. Diffusion is also essential for cell proliferation. These processes have been of interest in life, biomedical and biochemical sciences (see examples in Refs. \cite{cellpaper1,cellpaper2}).

Physically, the origin of this slowing down due to smaller $\hbar$ is related to the decrease of the Bohr radius \eqref{bohr} as the classical regime with smaller $\hbar$ is approached. This results in the increase of the cohesive energy in Eq. \eqref{rydberg} via Eq. \eqref{direct}, making it harder to flow and diffuse.

Large viscosity increase (think of viscosity of tar or higher) may mean that life might not exist in its current form or not exist at all. One might hope that cells could still survive in such a Universe by finding a hotter place where overly viscous and bio-unfriendly water is thinned. This would not help though: $\eta_{min}$ sets the minimum below which viscosity can not fall regardless of temperature or pressure. This applies to any liquid and not just water and therefore to all life forms using the liquid state to function.


We therefore see that water and life are well attuned to the degree of quantumness of the physical world (in conjunction with other fundamental constants and parameters). The same applies to other fundamental constants in Eq. \eqref{etamin} such as $e$, $m_e$ and to a smaller degree to $m_p$.

The results in this Chapter add another layer to the discussion of the anthropic principle, sometimes referred to as the anthropic argument \cite{adamsreview} or anthropic observation \cite{smolin}. Eliciting different views \cite{barrow,barrow1,vilenkin,hoganreview,adamsreview,uzanreview,carr,carr1,smolin}, this term is a collection of related ways to rationalise the observed values of fundamental constants by proposing that these constants serve to create conditions for an observer to emerge and hence are not unexpected. Developing this argument often involves an ensemble of disjoint universes and a physical mechanism to generate this ensemble. Then, a relatively small number of universes have the right values of fundamental constants, and we find ourselves in one of those universes and measure those constants. Alternatives include introducing the natural selection argument in cosmology, explaining the observed values of fundamental constants \cite{smolin}.

In these discussions, there are several types of conditions that need to be met for life and observers to exist. These conditions involve the range of effects, starting from cosmological processes and ending with nuclear synthesis discussed in the Introduction. Nuclear reactions are high-energy processes. Condensed matter physics involves much lower energies, and our earlier discussion showed how fundamental constants govern water viscosity. This adds a biological and biochemical aspect to the discussion of the anthropic principle. We can ask what change of water viscosity and diffusion constant from their current values is needed to disable cellular and inter-cellular biological processes essential to life. For example, this can happen if water became too viscous due to the lower viscosity bound getting larger, necessitating higher viscosity at all conditions. Once this is known, we can readily calculate the corresponding change of fundamental constants setting this lower bound using, for example, Eq. \eqref{etamin}.

One might think that the constraints on fundamental constants from star formation or nuclear synthesis are already tight enough to keep water viscosity from taking unwanted values not conducive to life. There are two points to consider here. First, it is possible to substantially change the lower bounds for kinematic and dynamic viscosity and at the same time keep the fine structure constant $\alpha=\frac{e^2}{\hbar c}$ and the electron-to-proton mass ratio $\beta=\frac{m_e}{m_p}$ intact, with no consequences for star formation and nuclear synthesis. 

Second, different effects involved in the existing hierarchy of observed fundamental constants and operating at different levels \cite{barrow,carr,carr1} have different tolerance to life-disabling variations \cite{hoganbook,adamsreview}. A small, compared to large, range of allowed fundamental constants is interesting because it tells us how special our Universe is and sets the weight of the anthropic argument. We have seen that sustaining liquid-based life (including water-based life) imposes constraints on fundamental physical constants which are additional to and different from what has been discussed before in nuclear synthesis. These constraints come from condensed matter physics and involve biology and biochemistry, adding a higher level to the hierarchy of life-enabling effects \cite{barrow,barrow1,carr,carr1}. It remains to be seen how tight these constraints are compared to constraints discussed in particle physics, astronomy and cosmology.


Exploring these and related issues further is important and invites an inter-disciplinary research. This interdisciplinarity has previously included some chemical and biochemical aspects of life \cite{carr-rees,barrow1,ellisuzan}, however the overall focus was on particle physics, astronomy and cosmology and on production of heavy elements in stars \cite{barrow,barrow1,hoganreview,adamsreview,uzanreview,carrbook,finebook,carr-rees}. On the other hand, fundamental insights from condensed matter physics were not explored, and this overview illustrates the benefits of this consideration.

It is useful to note that testability and falsifiability of a physical model involved in the current discussions of the anthropic principle is a central issue \cite{smolin}. On the other hand, the physical model underlying the viscosity minima comes from condensed matter physics with plentiful opportunities to test and falsify it. As we have seen earlier, the physical model underlying viscosity minima benefits from agreeing with a wide range of experimental data.


\subsection{Quantum liquids}

Quantum liquids are liquids where the effects of quantum statistics, Fermi or Bose, become operative at low temperature on the order of $\sim$ 1 K. Quantum liquids is a large area of research with long history where superfluidity in liquid helium plays an important role \cite{landaustat,pinesnoz}.

Despite this long history, some central problems remain not understood. Pines and Nozieres observe \cite{pinesnoz} that ``microscopic theory does not at present provide a quantitative description of liquid He II'' (``II'' here refers to helium below the superfluid transition temperature of about 2.2 K). This is in contrast to superconductivity where superconducting properties emerge from a microscopic Hamiltonian. For quantum fluids, a microscopic theory exists only for models of dilute gases or models with weak interactions where perturbation theory applies such as the Bogoliubov theory. Griffin broadly agrees with the assessment of Pines and Nozieres and says that we can't make quantitative predictions of superfluid $^4$He on the basis of existing theories and depend on experimental data for guidance \cite{griffin}. Interestingly, Griffin attributes the theoretical problems of understanding the superfluid He to the ``difficulties of dealing with a liquid, whether Bose-condensed or not''. In other words, he recognizes that the general problems of liquid theory discussed in Chapter \ref{minimalv}: the no-small parameter problem related to the combination of strong interactions and dynamical disorder.

Compared to several decades ago, research into liquid helium superfluidity has been slowing down. We have a set of important results and we know that several fundamental problems remain unresolved but we don't know where the next important insight is likely to come from. One insight we learned from classical liquids is that considering microscopic details of their dynamics and the combined oscillatory and diffusive components of particle motion in particular is the key to understanding liquids. This motion governs collective excitations, phonons, in liquids which, in turn, govern liquid thermodynamic properties \cite{ropp}. It may well be that these dynamical details will similarly need to be incorporated in the future microscopic theory of liquid helium.

In Chapter \ref{minimalv}, we used the microscopic mechanism of molecular motion in liquids to derive lower bound of liquid viscosity. In view of the need for microscopic theory of liquid helium, it is interesting to see whether we can discuss the minima of He viscosity on the basis of the same molecular mechanism as in classical liquids.

\begin{figure}
\begin{center}
{\scalebox{0.35}{\includegraphics{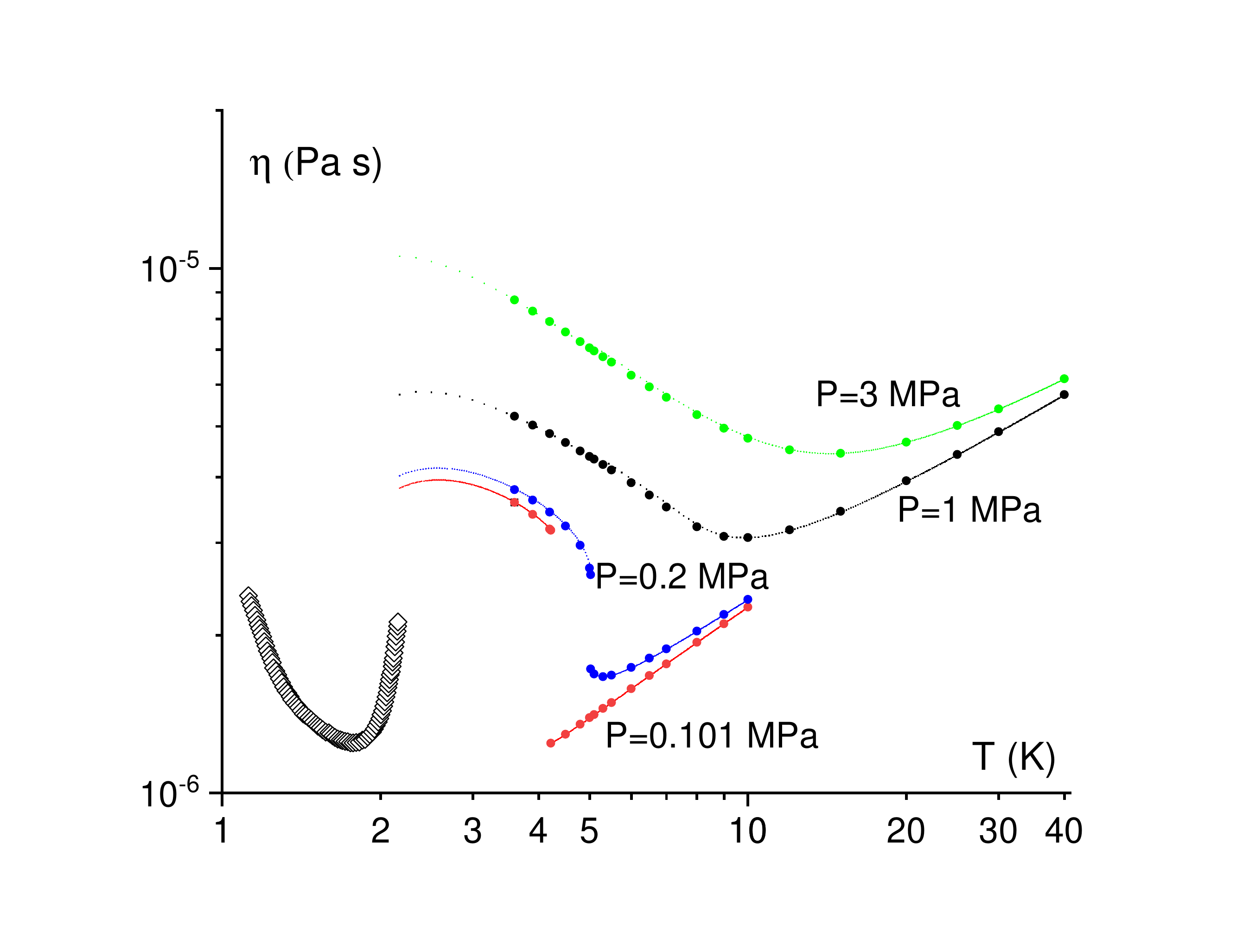}}}
{\scalebox{0.35}{\includegraphics{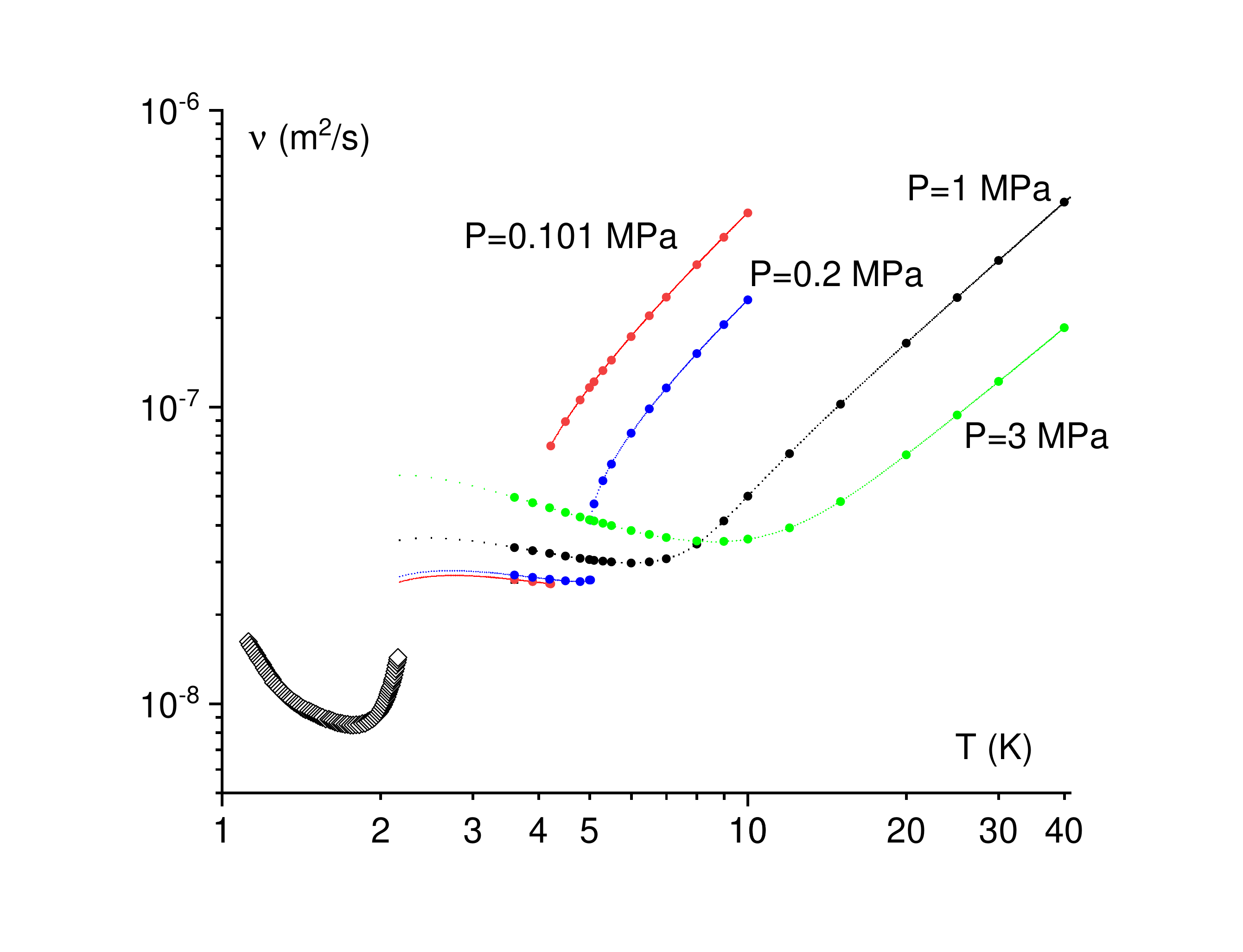}}}
\end{center}
\caption{Dynamic (top) and kinematic (bottom) viscosity of liquid $^4$He above and below the $\lambda$-point at about 2.2 K. Dotted lines are from Ref. \cite{nist}. Solid circles are from Ref. \cite{nist1}. Open diamonds correspond to viscosity of He II at atmospheric pressure from Ref. \cite{hevisc}.
}
\label{quvisc}
\end{figure}

In Figure \ref{quvisc}, we have compiled several sets of experimental data related to liquid $^4$He. The data represented by finely-spaced points and lines above the superfluid transition temperature $T_c=2.17$ K ($\lambda$-point) are from NIST \cite{nist}. These plots include interpolation artefacts at low temperature. These are usually unimportant in a wider temperature range, however here we are interested in liquid helium which exists in a narrow temperature range. For this reason, we also show the experimental points on which the NIST curves are based \cite{nist1} as bullet points. We also show viscosity of He II below $T_c$, attributed to the normal component with nonzero viscosity \cite{hevisc}. The kinematic viscosity is calculated using density from Ref. \cite{nist1}.

There are several observations from Figure \ref{quvisc}. First, we observe that viscosity have minima in both phases of He, He I and He II, similarly to other classical liquids in Figure \ref{visc}. Similar viscosity minima are also seen in $^3$He \cite{he3visc}.

We next observe that in He I above the critical pressure ($P_c=0.23$ MPa), viscosity minima in Figure \ref{quvisc} (especially the minima of kinematic viscosity) are not far from those seen in Figure \ref{visc} in classical liquids. The minima are somewhat lower in helium because (a) the pressure in Fig. \ref{quvisc} is lower and (b) viscosity minima in helium are understandably lower than in other liquids because the inter-particle interactions in He are particularly weak. Similarly to Figure \ref{visc}, viscosity minima increase with pressure. Second, viscosity has a jump due to the liquid-gas transition in the subcritical regime. The jump starts close to the viscosity minima in the supercritical regime at 1 and 3 MPa. The viscosity still have minima, albeit these are not smooth as in the supercritical state. This behavior is similar to that in Figure \ref{3}b. Third, although the viscosity minimum of He II is lower than in the liquid He I at atmospheric pressure by about a factor of 2-3 for both dynamic and kinematic viscosity in Fig. \ref{quvisc} (this can be attributed to He II considered to be a mixture of normal and superfluid components, see Refs. \cite{tisza,landaumigdal,balibar2,balibar,pinesnoz} for original papers and reviews), it is of the same order of magnitude.

Earlier in this Chapter, we showed that viscosity minima in classical liquids are set by fundamental physical constants. The similarity between viscosity minima in He I and He II in Figure \ref{quvisc} suggests that the minimum in He II is similarly set by these constants. This, in turn, indicates that the {\it mechanism} setting the viscosity minimum in He II may be similar. Indeed, the calculation of the viscosity minima in terms of fundamental constants in Section \ref{vminima} is based on a particular regime of particle dynamics corresponding to the crossover between liquidlike and gaslike regimes. The closeness of calculated and observed viscosity minima is therefore informative in a sense of microscopic dynamics: viscosity decreasing with temperature is related to the combined oscillatory and diffusive particle motion, whereas viscosity increasing with temperature is indicative of purely diffusive motion. This picture is consistent with path-integral simulations of He: the minima of velocity autocorrelation function, associated with the liquid-like combined oscillatory and diffusive motion \cite{Brazhkin2013}, are seen at 1.2 K \cite{nakayamapnas} where viscosity decreases in Figure \ref{quvisc}.

Clearly more work is needed to ascertain the nature of microscopic motion in liquid He and its relation to observed properties including superfluidity. Here, we see how the discussion of viscosity minima, their origin and value in terms of fundamental constants has the potential to provide interesting insights into microscopic dynamics in quantum liquids. This is important in view of constructing a microscopic theory of He II. Earlier in this section, we quoted the observation of Pines and Nozieres of the absence of a microscopic theory of He II. Such a theory would have to incorporate the microscopic particle of dynamics in liquid helium, and viscosity minima provide an insight into this dynamics.

Previously, the behavior of helium viscosity was discussed in terms unrelated to microscopic dynamics of particles. Landau and Khalatnikov calculated viscosity due to scattering of phonons and rotons by each other, with the result that viscosity decreases with temperature \cite{landauvisc}. This includes a provision that this result does not hold in the range where viscosity increases with temperature because of the proximity of the $\lambda$-point. Tisza, on the other hand, considered part of temperature range where viscosity increases with temperature, and attributes it to the gas-like behavior described by the gas kinetic theory. This was done in one of Tisza's pioneering papers \cite{tisza} introducing the two-fluid model of liquid helium (see Refs. \cite{balibar2,balibar,pinesnoz} for review of the two-fluid model). Dash \cite{dash} considers the entire regime where viscosity first decreases, goes through the minimum and then increases as in Figure \ref{quvisc} and explains this non-monotonic behavior by combining the Landau and Khalatnikov model with the model where viscosity increases due to the increasing normal fluid fraction.

\subsection{Quark-gluon plasma}
\label{qgp}

Differently from condensed matter systems, the subject of this review, the quark-gluon plasma (QGP) \cite{qgp} is a high-energy system. It is nevertheless interesting to mention the QGP here, for two reasons. First, a bound for viscosity-related property was proposed for the QGP. Second, the kinematic viscosity of the QGP is remarkably close to the viscosity minima discussed in Section \ref{vminima}.

In Section \ref{minimum}, we mentioned fundamental problems involved in liquid theory due to strong interactions. The same problem exists in strongly-coupled field theories where the perturbation theory does not apply. In some cases, it is possible to derive closed results using the duality between strongly-coupled field theories and weak gravity duals (see, e.g. Rev. \cite{malda} for review). Using this approach, the lower bound for viscosity-to-specific entropy ratio was derived as \cite{kss}:

\begin{equation}
\frac{\eta}{s}\ge\frac{\hbar}{4\pi}
\label{kss}
\end{equation}

\noindent where $s$ is the volume density of entropy.

The bound \eqref{kss} was referred to as the ``perfect fluidity'' and is being explored in different systems, including strongly-interacting Bose liquids, ultracold Fermi gases and quark-gluon plasma \cite{schafer}. This extends to the viscosity of quasiparticles in graphene \cite{grapheneta}. Later work considered how this and other bounds can be understood in the picture involving the Planckian relaxation time

\begin{equation}
\tau_{\rm P}=\frac{\hbar}{T}
\end{equation}

\noindent and related these and connected ideas to condensed matter systems including electron and spin transport properties \cite{behnia,zaanen,hartnoll,hartnoll1,spin,bruin,hbarm4,legros,sachdev,nussinov1} as well superconductivity and superfluidity \cite{volovik}.

$\tau_{\rm P}$ sets the limiting value of the relaxation time at a given temperature. In this sense, it is different from other bounds discussed in this review which are independent of external parameters and are set by fundamental physical constants.

The viscosity of the QGP has been measured experimentally: $\eta=5\cdot 10^{11}$ Pa$\cdot$s \cite{schafer}. Although $\eta$ is about 15 orders of magnitude larger than the viscosity of water at room conditions, the kinematic viscosity of the QGP is \cite{qgp}

\begin{equation}
\nu_{\rm QGP}^{exp}\approx 10^{-7} \frac{{\rm m}^2}{\rm s}
\end{equation}

\noindent and is close to the viscosity minima of ordinary liquids in Figure \ref{visc}b as well as fundamental viscosity in Eq. \eqref{nuf}.

This similarity is remarkable, given the 15 orders of magnitude difference in $\eta$ and that the two systems have disparate interactions and fundamental theories. A hint for this remarkable similarity comes from the universality of the dynamical crossover discussed in Section \ref{minimum}. At the crossover, particle dynamics is neither liquidlike with many oscillations and occasional jumps nor gaslike where $L\gg a$, but instead is at the border between the two regimes. At this border, kinematic viscosity turns out to be fixed by the fundamental constants only and independent of charge as mentioned in Section \ref{vminima}. This can help explain the similarity of $\nu_{min}$ of ordinary liquids and the quark-gluon plasma \cite{qgp}. This also suggests that the QGP may be close to the dynamical crossover in the sense discussed in Section \ref{minimum}.

The similarity of $\nu$ between the QGP and liquids at the minimum interestingly suggests that the flow properties of these disparate systems is similar. This is seen from the Navier-Stokes equation \eqref{navier} or its relativistic analogue \cite{qgp}.

The lower bound of the ratio \eqref{kss} was interestingly compared to real liquids such as N$_2$ and H$_2$O and found to be about 25 times smaller than viscosity minima in liquids. Most of this difference can be understood on the basis of elementary viscosity $\iota$ \eqref{iota1} which serves as an analogue of $\frac{\eta}{s}$ in \eqref{kss} because $\iota$ is the ratio of viscosity and number density $\frac{1}{a^3}$. The origin of this difference is the presence of the factor $\left({\frac{m}{m_e}}\right)^{\frac{1}{2}}$ in Eq. \eqref{iota1} \cite{sciadv1}. This factor is specific to condensed matter and does not feature in Eq. \eqref{kss} derived from a theory based on holographic correspondence and string theory.

\subsection{What is ``fundamental''?}
\label{weinbergfun}

In this Chapter, we have discussed bounds to viscosity set by fundamental physical constants. There is a truly fascinating history of earlier and ongoing effort to understand the origin and rationalise the values of fundamental constants including the dimensionless ones such as the fine structure constant $\alpha$, proton-to-electron mass ratio $\frac{m_e}{m_p}$ and so on \cite{barrow,uzanreview}. A possibility was raised that the fundamental constants might not even be fixed and vary in different epochs \cite{gamow}. Understanding fundamental constants, if feasible, is probably one of the ultimate grand challenges in physics.

Commenting on prospects to understand fundamental constants, Weinberg observes that the membership of fundamental constants depends on a theory or effects considered \cite{weinberg}. Viscosity of water serves as ``fundamental'' in hydrodynamics, whereas electron mass and electron charge play that function in atomic physics. There are perhaps two senses in which the term ``fundamental'' is discussed here. First, the hydrodynamic theory makes predictions about liquid flow and involves viscosity as a pre-determined parameter whose calculation can not be done and is not required in the hydrodynamic theory itself. Second, we can ask whether this or other similar parameter can be calculated on the basis of another, more fundamental, underlying theory. There is currently a limit to how fundamental we can go: calculating fundamental physical constants can not currently be done not because the calculation is too complicated (as for the viscosity of water, notes Weinberg), but because we don't know of anything more fundamental. On the other hand, condensed matter physics should in principle be able to provide tools to calculate water viscosity, although this remained very hard in view of general issues involved in liquid theory and viscosity in particular as discussed in Section \ref{minimum}.

The results in this Chapter suggest that despite difficulties involved in calculating viscosity as a ``fundamental'' parameter in fluid mechanics, viscosity is nevertheless governed by true fundamental physical constants (see Eq. \eqref{nuf}). These constants set bounds for viscosity and its values in a fairly wide range of parameters on the phase diagram.

\section{Thermal conductivity}
\label{thermal}

\subsection{Thermal conductivity of insulators and dynamical crossover}
\label{thermal1}

In this Section, we consider a property different to viscosity: the ability to conduct heat. We consider insulating systems where the conductivity is due to ions. In Chapter \ref{electronic}, we discuss thermal conductivity by electrons.

Thermal energy can be carried by phonons and electron quasi-particles in solids and liquids or molecular collisions in gases \cite{ashcroft,chapman}. Although these two mechanisms of heat transfer, by collective excitations or particles, are conceptually simple, they can interestingly interact with other processes and give rise to a rich variety of effects. These effects are currently explored in a variety of materials including insulators, strange metals and cuprate superconductors, where new mechanisms are invoked to explain the experimental data (see, e.g., Refs. \cite{zaanen,hartnoll1,behnia,bruin}). This involves bounds on thermal conductivity based on uncertainty relations and often involve temperature-dependent Planckian relaxation time $\tau_{\rm P}=\frac{\hbar}{T}$ mentioned in Section \ref{qgp}.

Thermal conductivity $\kappa$ is defined as the proportionality coefficient between the heat current density and the temperature gradient (e.g., $J_x=\kappa_{xx}\frac{\partial T}{\partial x}$ in the $x$-direction). The propagation of heat is given by the heat equation

\begin{equation}
\frac{\partial T}{\partial t}= \alpha\frac{\partial^2T}{\partial x^2}
\label{Heat}
\end{equation}

\noindent where $\alpha=\frac{\kappa}{\rho c_p}$ is thermal diffusivity, $\rho$ is density and $c_p$ is heat capacity per mass unit.

Eq. \eqref{Heat} is analogous to the Navier-Stokes equation \eqref{navier}. Similarly to the kinematic viscosity governing flow in Eq. \eqref{navier}, $\alpha$ quantifies the propagation of thermal energy.

Similarly to viscosity, the heat transport coefficients $\kappa$ and $\alpha$ vary in a wide range and depends strongly on the system, temperature and pressure. Yet we will see below that the lower bound of these properties is identical to that of viscosity.

\begin{figure}
\begin{center}
{\scalebox{0.37}{\includegraphics{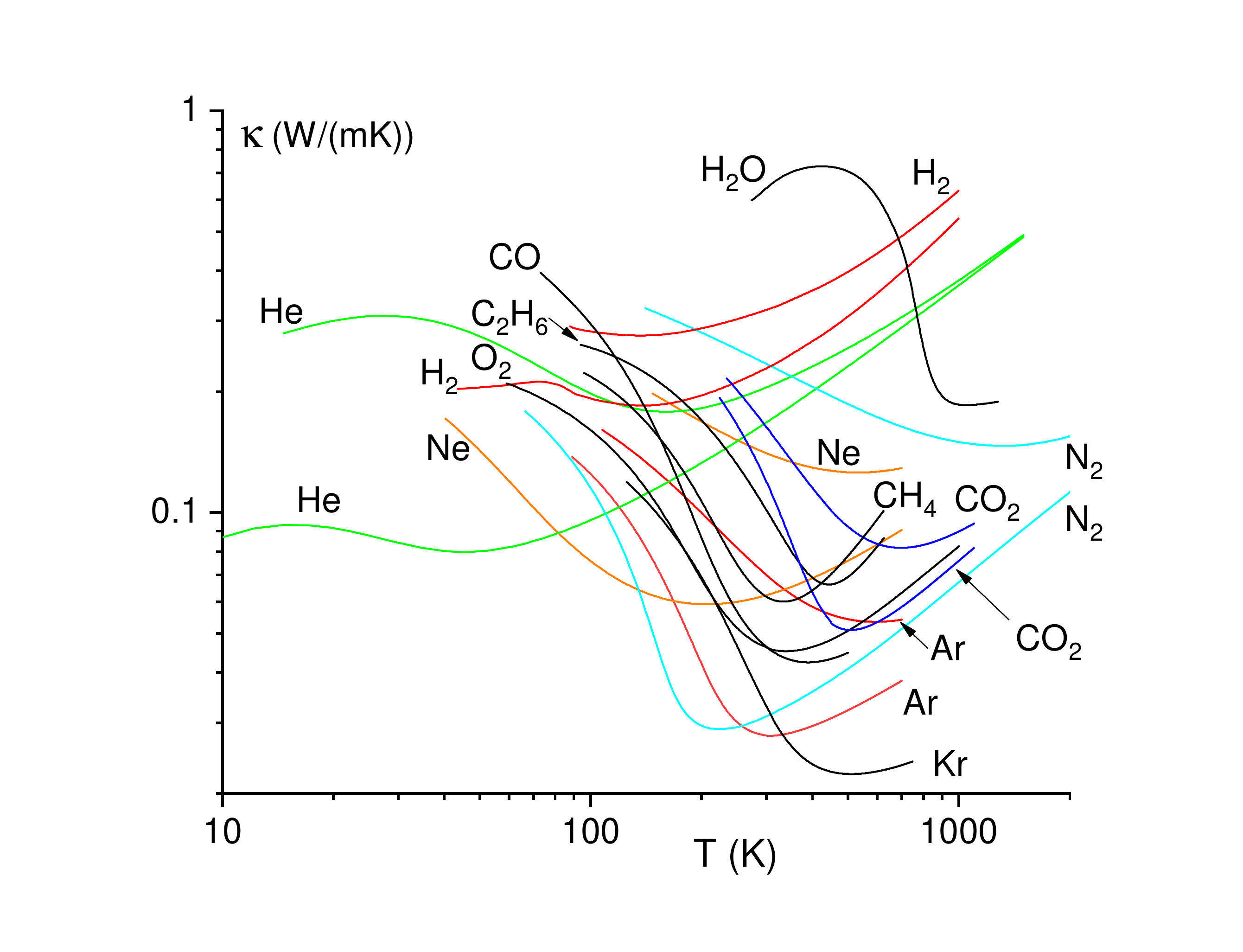}}}
{\scalebox{0.37}{\includegraphics{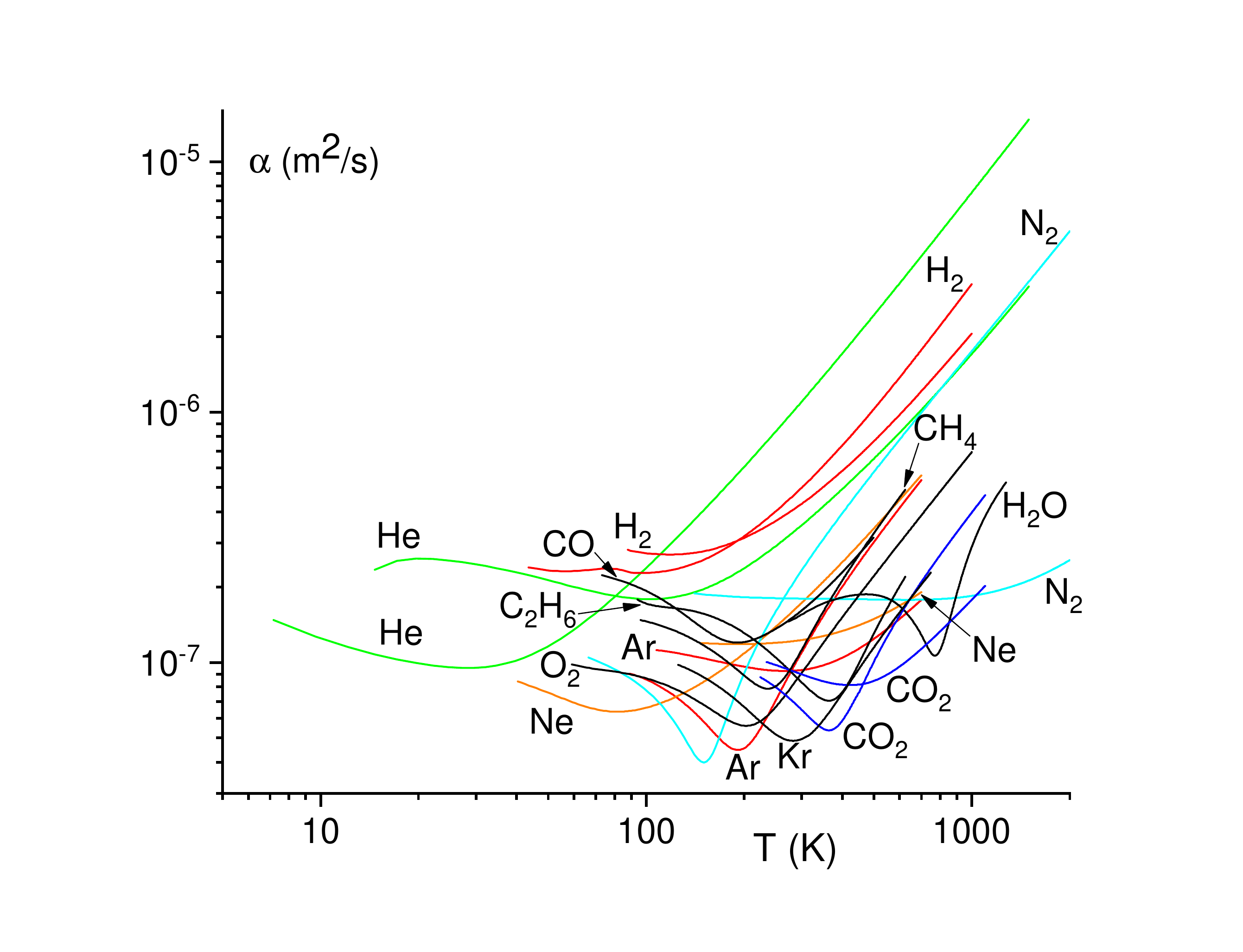}}}
\end{center}
\caption{Experimental thermal conductivity $\kappa$ (top) and thermal diffusivity $\alpha$ (bottom) of noble, molecular and network liquids \cite{nist} showing minima. $\kappa$ and $\alpha$ for Kr, O$_2$, H$_2$O, CH$_4$, C$_2$H$_6$ and CO are shown for pressure $P=30$ MPa, 30 MPa, 70 MPa, 20 MPa, 20 MPa and 20 MPa, respectively. $\eta$ for Ar, Ne, He, N$_2$, H$_2$ and CO$_2$ are shown at two pressures each: 20 and 100 MPa for Ar, 50 and 300 MPa for Ne, 20 and 100 MPa for He, 10 MPa and 500 MPa for N$_2$, 50 MPa and 100 MPa for H$_2$, and 30 and 90 MPa for CO$_2$. The minimum at higher pressure is above the minimum at lower pressure for each fluid.}
\label{thermalfig}
\end{figure}

We have collected available experimental data \cite{nist} of $\kappa$ in several noble (Ar, Ne, He and Kr), molecular (N$_2$, H$_2$, O$_2$, CO$_2$, CH$_4$ C$_2$H$_6$ and CO) and network fluids (H$_2$O). This selection includes industrially important supercritical fluids such as CO$_2$ and H$_2$O. We have calculated $\alpha=\frac{\kappa}{c\rho}$ using the experimental values of $c_p$ and $\rho$ and show both $\kappa$ and $\alpha$ in Figure \ref{thermalfig}. For some fluids, we show the data at two different pressures. As in the case of viscosity in Figure \ref{visc}, the low pressure was chosen to be sufficiently far above the critical pressure so that the data are not affected by near-critical anomalies. The high pressure was chosen to (a) make the pressure range as wide as possible and (b) be low enough in order to see the minima in the available temperature range.

We observe that $\kappa$ and $\alpha$ universally have minima, similarly to viscosity in Figure \ref{visc}. We also observe that $\kappa$ can have maxima at low temperature related to the competition between the increase of heat capacity due to phonon excitations in the low-temperature quantum regime and decrease of the phonon mean free path $l$ as in solids. In H$_2$O, the broad maximum is related to water-specific anomalies including broad structural transformation between differently-coordinated states.

We now discuss the reason why $\kappa$ and $\alpha$ have minima in Figure \ref{thermalfig}. In solids and systems where heat is carried by phonons, the thermal conductivity $\kappa$ is $\kappa=cvl$, where $c$ is the specific heat per volume unit \cite{ashcroft}, $v$ is the speed of sound, $l$ is the phonon mean free path and we dropped the numerical factor on the order of unity. Then, thermal diffusivity $\alpha$ is

\begin{equation}
\alpha=vl
\label{t1}
\end{equation}

In gases, $\alpha$ can be written in the same way as (\ref{t1}), but - and this reflects the difference between heat transfer in solids and gases - $v$ in (\ref{t1}) corresponds to the average velocity of gas molecules and $l$ to the molecule free path \cite{chapman}.

We can now see that the minimum of $\alpha$ is due to the dynamical crossover between the liquid-like and gas-like regimes of particles dynamics discussed in Section \ref{minimum}. The liquid phonon states consist of one longitudinal mode and two transverse modes propagating above the threshold value in $\omega$ or $k$-space \cite{ropp}. Temperature increase has two effects on $\alpha$ in Eq. (\ref{t1}): both the phonon mean free path $l$ and the speed of sound $v$ decrease. However, the decrease of $v$ and $l$ can not continue indefinitely: $l$ is limited by either the phonon wavelength \cite{slack} or its shortest value comparable to the interatomic separation $a$ (see the discussion of the reduction of $l$ close to $a$ by Kittel \cite{kittelmeanf} in disordered glasses). Similarly, $v$ decreases with temperature at the dynamical crossover discussed in Section \ref{minimum} where it becomes comparable to the particle thermal speed at the Frenkel line, $v_t$. At this crossover, the oscillatory component of molecular motion in liquids is lost, and molecules start moving in a purely diffusive manner. In this regime, $l$ becomes the particle mean free path, $l_p$. $l_p$ and $v_t$ both increase with temperature. Therefore, $\alpha$ in Eq.(\ref{t1}) has a minimum.

The same mechanism leading to a minimum applies to $\kappa=c\rho\alpha$. $\rho$ and $c$ monotonically decrease with temperature \cite{ropp}, hence the minima of $\alpha$ and $\kappa$ can take place at somewhat different temperature.

Before evaluating $\alpha_{min}$, let us see how well we can estimate $\kappa$ at the minimum, $\kappa_{min}$. The speed of sound $v$ in the Debye model is $v=\frac{a}{\tau_{\rm D}}$ (at the crossover where $\tau$ becomes comparable to the time it takes the molecule to move distance $a$ and where $\tau\approx\tau_{\rm D}$ as discussed above, $v$ becomes approximately equal to thermal velocity). Recalling that $c$ featuring in $\kappa=cvl$ is the temperature derivative of energy density \cite{ashcroft}, $c=\frac{c_v}{a^3}$, where $c_v$ is heat capacity per atom at constant volume (if the derivative is taken at constant volume) and $a^{-3}$ is the concentration. At the minimum corresponding to the dynamical crossover at the Frenkel line, $c_v$ is close to $2$, reflecting the disappearance of two transverse modes \cite{ropp}. Setting $l=a$, $v=\frac{a}{\tau_{\rm D}}=\frac{1}{2\pi}\omega_{\rm D}a$, where $\omega_{\rm D}$ is Debye frequency, gives

\begin{equation}
\kappa_{min}=\frac{1}{\pi}\frac{\omega_{\rm D}}{a}
\label{km}
\end{equation}

Taking the typical values of $a=$3-6 \AA\, $\frac{\omega_{\rm D}}{2\pi}$ on the order of 1 THz and reinstating $k_{\rm B}$, we find $\kappa_{min}$ in the range $0.05-0.09$ $\frac{{\rm W}}{{\rm mK}}$. This is consistent with typical values seen in Fig. \ref{thermalfig}a. We also observe that high pressure reduces $a$ and increases $\omega_{\rm D}$ in Eq. \eqref{km}, hence we predict that $\kappa_{min}$ increases with pressure as a result. This is in agreement with the experimental behavior in Fig. \ref{thermalfig}. Another prediction of Eq. \eqref{km} is the reduction of $\kappa_{min}$ with mass $m$ due to $\omega_{\rm D}\propto\frac{1}{\sqrt{m}}$. Consistent with this prediction, $\kappa_{min}$ in Fig. \ref{thermalfig}a tend to be lower for heavier systems such as Kr.

We note that the minima of $\kappa$ of most liquids in Fig. \ref{thermalfig}a are still lower than thermal conductivity in low-$\kappa$ solids such as SnSe ($\kappa=0.23~\frac{{\rm W}}{{\rm mK}}$) where it is considered to be exceptionally low \cite{snse}. For Kr, $\kappa_{min}$ is about 10 smaller than the ultralow value of $\kappa$ in SnSe.

\subsection{Lower bound on thermal diffusivity}

We now evaluate $\alpha$ at its minimum, $\alpha_{min}$. As discussed above, $l\approx a$ at the minimum at the dynamical crossover. Using $v=\frac{a}{\tau_{\rm D}}=\frac{1}{2\pi}a\omega_{\rm D}$ in Eq. \eqref{t1} as before gives

\begin{equation}
\alpha_{min}=\frac{1}{2\pi}\omega_{\rm D}a^2
\label{alpham}
\end{equation}

Eq. \eqref{alpham} is the same as $\nu_{min}$ in Eq. \eqref{nu} in Section \ref{vminima}. Therefore, repeating the same steps as in Section \ref{vminima}, we find \cite{prbthermal}:

\begin{equation}
\alpha_{min}=\nu_{min}=\frac{1}{4\pi}\frac{\hbar}{\sqrt{m_em}}
\label{alpham1}
\end{equation}

\noindent giving rise to the fundamental thermal diffusivity $\alpha_f$ as in Eq. \eqref{nuf}:

\begin{equation}
\alpha_f=\frac{1}{4\pi}\frac{\hbar}{\sqrt{m_em_p}}\approx 10^{-7} \frac{\rm{m}^2}{\rm{s}}
\label{alphaf}
\end{equation}

Eq. \eqref{alphaf} is consistent with the experimental results in Figure \ref{thermalfig}b. Similarly to viscosity discussed in Chapter \ref{minimalv}, this shows how fundamental constants set the characteristic scale of physical properties including complicated ones such as thermal conductivity and diffusivity.

The prediction of Eq. \eqref{alpham1} can be compared to experiments. In Table \ref{tab1} we compare $\alpha_{min}$ calculated according to (\ref{alpham1}) to the experimental $\nu_{min}$ \cite{nist} for all liquids shown in Fig. \ref{thermalfig}. The ratio between experimental and predicted $\alpha_{min}$ is in the range of about $0.9-4$. The ratio is the largest for fluids under high pressure (e.g. N$_2$ at 500 MPa and Ar at 100 MPa) which Eq. (\ref{nu1}) does not account for. For the lightest liquid, H$_2$, experimental $\alpha_{min}$ is close to the theoretical fundamental thermal diffusivity viscosity (\ref{alpham}). We therefore find that (\ref{nu1}) is consistent with the experimental data, with caveats discussed in Section \ref{vminima} related to approximations involved.

\begin{table}[ht]
\begin{tabular}{ l| c c c c}
                   & $\alpha^{th}_m=\nu^{th}_m$     & $\alpha^{exp}_m$       & $\nu^{exp}_m$ &$\nu_{min}/\alpha_{min}$\\
\hline\\
Ar (20 MPa)        & 3.4                   & 4.5                & 5.9  & 1.3\\
Ar (100 MPa)       & 3.4                   & 9.3                & 7.7  & 0.8\\
Ne (50 MPa)        & 4.8                   & 6.4                & 4.6  & 0.7\\
Ne (300 MPa)       & 4.8                   & 11.9               & 6.5  & 0.6\\
He (20 MPa)        & 10.7                  & 9.5                & 5.2  & 0.6\\
He (100 MPa)       & 10.7                  & 17.9               & 7.5  & 0.4\\
Kr (30 MPa)        & 2.3                   & 4.9                & 5.2 &1.1\\
N$_2$ (10 MPa)     & 4.1                   & 4.0                & 6.5  &1.6\\
N$_2$ (500 MPa)    & 4.1                   & 17.8               & 12.7 & 0.7\\
H$_2$ (50 MPa)     & 15.2                  & 22.8               & 16.3 & 0.7\\
H$_2$ (100 MPa)    & 15.2                  & 27.0               & 19.4 &0.7\\
O$_2$ (30 MPa)     & 3.8                   & 5.6                & 7.4  &1.3\\
H$_2$O (70 MPa)    & 5.1                   & 10.7               & 11.9 &1.1\\
CO$_2$ (30 MPa)    & 3.2                   & 5.4                & 8.0  &1.5\\
CO$_2$ (90 MPa)    & 3.2                   & 8.1                & 9.3  &1.2\\
CH$_4$ (20 MPa)    & 5.4                   & 7.9                & 11.0 &1.4\\
C$_2$H$_6$ (20 MPa)& 3.9                   & 7.0                & 12.0 &1.7\\
CO (20 MPa)        & 4.1                   & 12.0               & 7.7  &0.6\\
\end{tabular}
\caption{Theoretical (th) and experimental (exp) values for the thermal diffusivity $\alpha_{min}$ and the kinematic viscosity $\nu_{min}$ at the minima. All the quantities are displayed in units of $\times$10$^8$ m$^2$/s except from the last ratio which is dimensionless.}
\label{tab1}
\end{table}

The closeness of the minima of both properties, $\alpha_{min}$ and $\nu_{min}$ in Eq. \eqref{alphaf} is unexpected and surprising. Indeed, viscosity and thermal conductivity are physically distinct properties. They are measured in very different experiments. Yet Eq. \eqref{alphaf} predicts that their minima should be the same.

This prediction is checked in Table in \ref{tab1} where $\nu$ at the minima are calculated at the same pressure as $\alpha$. Consistent with the prediction of Eq. \eqref{alpham}, we observe that the experimental values of $\alpha_{min}$ and $\nu_{min}$ are close to each other. This agreement is also seen in the last column of Table \ref{tab1} where the ratio $\nu_{min}/\alpha_{min}$ is in the range 0.4-1.7.

We note that the temperatures of the minima of $\alpha_{min}$ and $\nu_{min}$ are somewhat different, nevertheless the closeness of $\alpha_{min}$ and $\nu_{min}$ implies that the Prandtl number, $\frac{\nu}{\alpha}$, is on the order of 1 at temperatures close to the minima. In other words, the transfer of energy and momentum takes place with the same velocity in this regime.

\begin{figure}[ht]
\begin{center}
{\scalebox{0.35}{\includegraphics{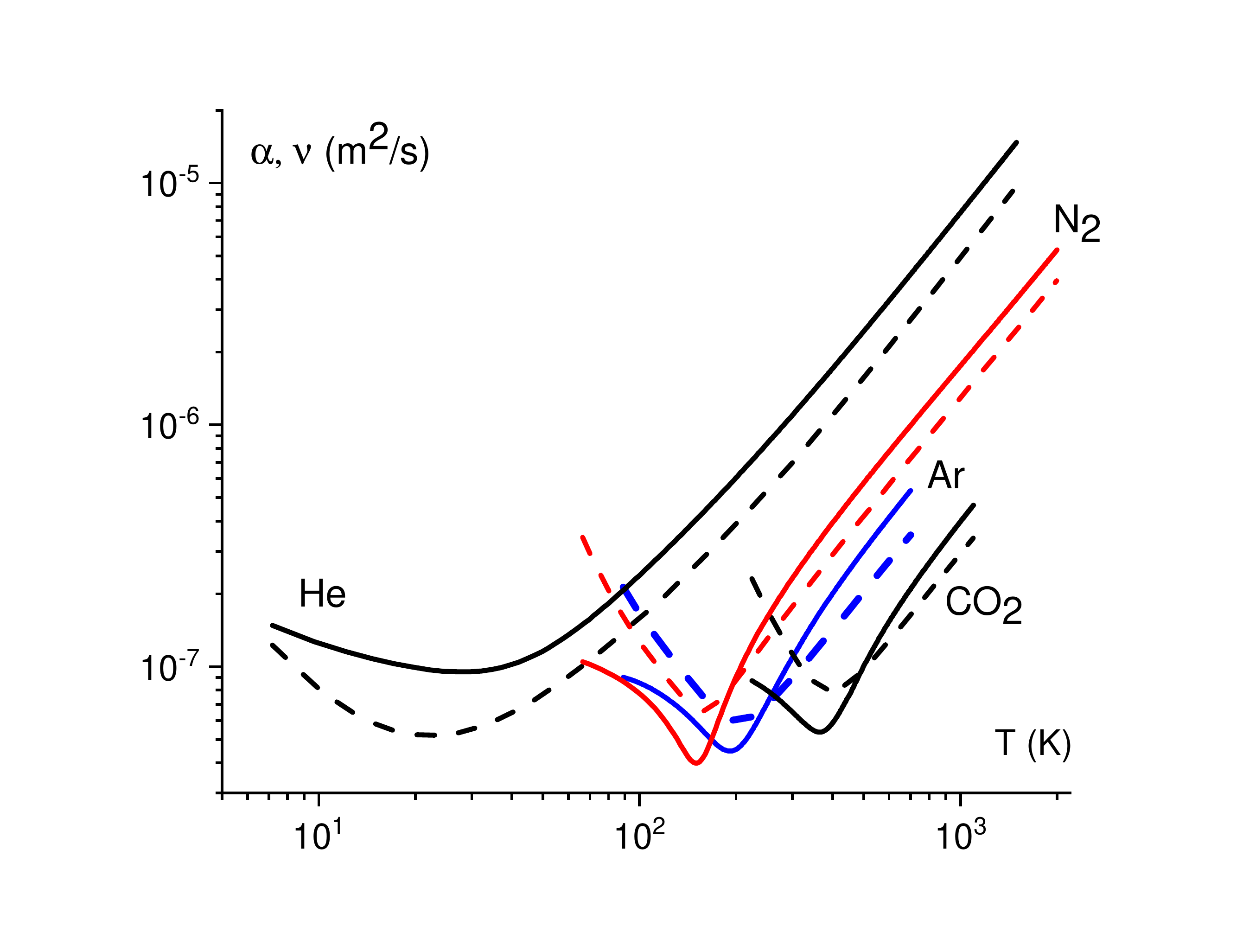}}}
\end{center}
\caption{Experimental thermal diffusivity $\alpha$ (solid lines) and kinematic viscosity $\nu$ (dashed lines) for He (20 MPa), N$_2$ (10 MPa), Ar (20 MPa) and CO$_2$ (30 MPa) \cite{nist}. Reproduced from Ref. \cite{prbthermal} with permission from the American Physical Society.}
\label{alphanu}
\end{figure}

To illustrate the closeness of $\alpha_{min}$ and $\nu_{min}$ further, we plot the experimental $\alpha$ and $\nu$ for two noble and two molecular liquids in Fig. \ref{alphanu} at the same pressures as in Fig. \ref{thermal} and observe the closeness of the minima of two properties.

Figure \ref{alphanu} prompts us to think about other interesting similarities as well as differences between kinematic viscosity and thermal diffusivity. We have already mentioned the first general similarity between $\alpha$ and $\nu$: they feature in the Navier-Stokes \eqref{navier} and heat equation \eqref{Heat} which have a similar form. Second, the dominant contribution to thermal conductivity in the low-temperature liquid-like regime is due to phonons as in solids. In the high-temperature gas-like regime, thermal conductivity is due to particle collisions. Viscosity, on the other hand, is due to the dynamics of individual particles and momentum they transfer in both liquid-like regime \eqref{v1} and gas-like regime \eqref{v2}. Therefore, viscosity and thermal conductivity are set by the same process at high temperature but by different processes at low temperature. Consistent with this prediction, Fig. \ref{alphanu} shows that temperature behavior of $\alpha$ and $\nu$ is more similar at high temperature as compared to low.

\subsection{Thermal diffusivity of electrons in metals}
\label{electronic}

The discussion of thermal conductivity in the previous section applies to systems where the dominant contribution is related to the motion of atoms, ions or molecules. We now consider systems where electron conductivity is important or dominant, such as metals.

Similarly to thermal diffusivity in insulating systems discussed in Section \ref{thermal1}, the minimum of electron thermal diffusivity, $\alpha_{min}^e$, corresponds to setting $l=a$ in Eq. (\ref{t1}): $\alpha_{min}^e=va$, where $v$ is the electron speed. This corresponds to the Ioffe-Regel crossover (see, e.g., Ref. \cite{iofferegel} and references therein). The electron velocity can be estimated as $v=\sqrt\frac{2E}{m_e}$, where $E$ is given by the Rydberg energy (\ref{rydberg}). Estimating $a$ as $a_{\rm B}$ in Eq. \eqref{bohr} as before give $\alpha_{min}^e$ as

\begin{equation}
\alpha_{min}^e=a_{\rm B}\left(\frac{2E_{\rm R}}{m_e}\right)^{\frac{1}{2}}
\label{alelec}
\end{equation}

Using $a_{\rm B}$ from Eq. \eqref{bohr} and $E_{\rm R}$ from Eq. \eqref{rydberg} in Eq. \eqref{alelec} gives

\begin{equation}
\alpha_{min}^e=\frac{\hbar}{m_e}\approx 10^{-4} \frac{\rm{m}^2}{\rm{s}}
\label{elec}
\end{equation}

Similarly to kinematic viscosity discussed in Section \ref{elementary}, $\alpha_{min}^{e}$ is consistent with an uncertainty relation applied to an electron located within a distance $a$, $m_eva\ge\hbar$.

Set by fundamental physical constants, the bound (\ref{elec}) is universal and does not depend on the system, in contrast to (\ref{alpham1}). Comparing to the fundamental thermal diffusivity due to ions in Eq. \eqref{alphaf}, we see that $\alpha_{min}^e$ is about $10^3$ times larger. This is due to smaller electron mass as compared to the proton mass.

We note that $\frac{\hbar}{m}$, where $m$ is the particle mass, has been discussed as the lower diffusivity bound for spin transport \cite{spin,hbarm1,hbarm2,hbarm3,hbarm4}.

Eq. \eqref{alelec} can be compared to the experimental data in liquid metals at high temperature where $l$ is expected to approach $a$. This would be an interesting analysis to perform.

\section{Minima on the phase diagram: theory and applications}
\label{minimaphase}

The discussion of fundamental limits on viscosity and thermal conductivity in the previous two Chapters were related to liquids and supercritical fluids. It turns out that these limits also enable us to discuss the limits of these properties in the entire phase diagram of matter.

Let us consider to what extent the minima of $\nu_{min}$ and $\alpha_{min}$ discussed for the liquid and supercritical states in Chapters \ref{minimalv} and \ref{thermal} apply to other parts of the phase diagram. The data showing the increase of $\nu$ with temperature in the gas regime in Fig. \ref{visc} is above the triple point. Below the triple point, $\nu$ is larger than $\nu_{min}$. This follows from observing that (a) $\nu$ increases above the sublimation line along the isobars and also increases along the sublimation line on lowering the temperature due to the exponential decrease of sublimation pressure \cite{landaustat} (we do not consider quantum effects), and (b) $\nu$ at the triple point is significantly larger than $\nu_{min}$ \cite{nist}. Therefore, $\nu_{min}$ corresponds to the minimum for both fluids and gases (phases where viscosity operates), including dilute low-temperature gases.

Considering now thermal diffusivity, we note that in solids $\alpha=vl$ is larger than $\alpha_{min}$, for two reasons. First, the speed of sound $v$ is faster. Second, the phonon mean free path $l$ is larger than that in liquids and is typically larger than $a$ at the UV cutoff. In gases, $\alpha=v_tl_p$, where $l_p$, the particle mean free path and $v_t$, thermal velocity, increase with temperature. At the minimum, $l\approx a$, and the speed of sound is approximately equal to the thermal speed of particles at the Frenkel line \cite{flreview}. Hence $\alpha=vl$ increases in both solids and gases, and the minimum of $\alpha$, $\alpha_{min}$, applies to all three states of matter.

We therefore see that $\alpha_{min}$ and $\nu_{min}$ represent minima on the phase diagram. However, $\alpha_{min}$ and $\nu_{min}$ behave differently in close proximity to the critical point. Indeed, viscosity diverges at the critical point \cite{xenonvisc}, and $\nu_{min}$ increases close to the critical point as a result. Therefore, $\nu_{min}$ gives the global minimum on the entire phase diagram. On the other hand, the isobaric heat capacity diverges much faster than $\kappa$ \cite{anisimov}. As a result, $\alpha$ at the critical point tends to zero. Therefore, $\alpha_{min}$ gives the minimum on the phase diagram of matter except in the close vicinity of the critical point.

As far as dynamic viscosity $\eta$ and thermal conductivity $\kappa$ are concerned, their minima operate in the liquid and supercritical parts of the phase diagram only, including the critical point where they increase. In the gas phase, $\eta$ and $\kappa$ can be arbitrarily small at low temperature. The same applies to $\kappa$ in solids where it tends to zero due to heat capacity becoming zero at low temperature and where the phonon mean free path saturates to a constant value set by either system size or scattering from defects.

Apart from ascertaining theoretical minima of the phase diagram, the fundamental limits of viscosity and thermal conductivity have practical implications. For example, designing a low-viscosity liquid in lubricating applications benefits from knowing that viscosity can not be lower than the fundamental bound. On the other hand, if viscosity is substantially larger than the bound, there is room for improvement which can be pursued. Similarly, low-viscosity and associated high diffusion is important in increasing deployment of supercritical fluids such as CO$_2$ and H$_2$O in cleaning, extracting and dissolving processes including environmental and green applications \cite{kiran2000,McHardy1998,Brunner2010,Alekseev2020,proctor2,flreview}. It was noted that improving the fundamental knowledge of the supercritical state properties is important for scaling up, widening, and increasing the reliability of these applications (see, e.g., Refs \cite{kiran2000,Leitner2002,Eckert1996,Savage1995,Beckman2004,Jessop1999}).

Similarly, the lower bound for thermal conductivity and diffusivity is informative when designing a system with superior thermal insulation properties. Small thermal conductivity is also important in other areas such as enhancing the thermoelectric effect. The figure of merit measuring the efficiency of thermoelectricity is inversely proportional to thermal conductivity. Therefore, the minimal thermal conductivity gives the maximal possible figure of merit, keeping all other factors unchanged. As mentioned in Section \ref{thermal1}, the exceptionally low thermal conductivity reported in Ref. \cite{snse} for the solid with high thermoelectric figure is still larger than theoretical lower bounds.

\section{Elastic properties}
\label{elastic}

\subsection{Elastic moduli}
\label{elastic1}

A convenient starting point of the discussion of elastic moduli is to write the system energy as \cite{aleksandr}

\begin{equation}
E=E_0f\left(\frac{V}{V_0}\right)
\label{elenergy}
\end{equation}

\noindent where $E_0$ and $V_0$ are energy and volume at zero temperature and pressure and $f$ is the function showing the energy dependence on volume.

Eq. \eqref{elenergy} gives the bulk modulus $K=V\frac{\partial^2 E}{\partial V^2}$ as

\begin{equation}
K=\frac{E_0}{V_0}\frac{V}{V_0}f^{''}\left(\frac{V}{V_0}\right)
\label{elenergy1}
\end{equation}

At zero pressure, the bulk modulus is

\begin{equation}
K_0=\frac{E_0}{V_0}f_0^{''}
\label{elenergy2}
\end{equation}

Early studies have showed that $f_0^{''}$ is close to 1 for systems with covalent bonding \cite{aleksandr}. Later work ascertained that $f_0^{''}$ is on the order of 1 for a wider class of solids, including metallic, molecular and noble systems \cite{brahemley,solozhen}. This implies that the bulk modulus and related elastic properties are largely governed by the
bonding energy density, or density of valence electrons:

\begin{equation}
K_0\approx\frac{E_b}{a^3}
\label{elenergy3}
\end{equation}

\noindent where $E_b$ is the bonding energy.

In diamond, this density is high due to fairly small ionic radius and four valence electrons \cite{brahemley,solozhen}. This gives diamond its uniquely large modulus of about 450 GPa.

If the bulk modulus is given by the density of cohesive energy, we can use Eq. \eqref{elenergy3} to estimate the ``fundamental bulk modulus'' $K_f$ in terms of fundamental physical constants as \cite{kirzh}

\begin{equation}
K_f=\frac{E_{\rm R}}{a_{\rm B}^3}\approx 147~{\rm Mbar}
\label{modfund}
\end{equation}

\noindent where $E_{\rm R}$ is the Rydberg energy in Eq. \eqref{rydberg} and $a_{\rm B}$ is the Bohr radius in Eq. \eqref{bohr}.

We note that elastic moduli have the dimension of pressure, and $K_f$ in Eq. \eqref{modfund} is often called the atomic pressure unit. At pressures above $K_f$, effects related to the overlap of inner electronic shells come into play, at which point solids metallise and become similar to each other and ultimately to the Thomas-Fermi plasma.

From Eqs. \eqref{rydberg} and \eqref{bohr}, we see that $K_f$ depends on fundamental physical constants $m_e$, $e$ and $\hbar$ as

\begin{equation}
K_f\propto\left(\frac{m_e^2e^5}{\hbar^4}\right)^2
\end{equation}

\noindent where we dropped numerical factors.

$K_f$ of about 147 Mbar represents an upper bound for the bulk modulus of condensed matter systems (solids and liquids) because $a_{\rm B}\approx 0.5$ \AA\ in \eqref{modfund} is smaller than the interatomic separation in real systems and the Rydberg energy corresponds to large cohesive energy.

As a rough estimate, we can see what Eq. \eqref{modfund} predicts for diamond where the interatomic separation is about 3 times larger than $a_{\rm B}$. According to Eq. \eqref{modfund}, this gives the bulk modulus smaller than $M_f$ by a factor of $3^3$, or about 540 GPa, close to that in diamond.

\subsection{Low-dimensional systems: surface tension and atomic force}
\label{lowd}

Similarly to elastic moduli, we can consider the surface tension as the surface energy density $\sigma=\frac{E}{r^2}$. Taking $E=E_{\rm R}$ and $r=a_{\rm B}$ as before gives

\begin{equation}
\sigma=\frac{E_{\rm R}}{a_{\rm B}^2}\approx 780~\frac{\rm N}{\rm m}
\label{sigma}
\end{equation}

\noindent or

\begin{equation}
\sigma\propto\frac{m_e^3e^8}{\hbar^6}
\end{equation}

\noindent in terms of fundamental constants.

Compared to the common surface tension in liquids, Eq. \eqref{sigma} gives a large value. For example, $\sigma$ in water and mercury bordering air is 0.07 and 0.5 $\frac{\rm N}{\rm m}$, respectively. The meaning of $\sigma$ in Eq. \eqref{sigma} is that gives maximally possible surface tension, and a fair comparison is with elastic moduli and mechanical properties of two-dimensional solids. The stiffest two-dimensional solid known, graphene, has $\sigma=340~\frac{\rm N}{\rm m}$ and the averaged breaking strength of 55~$\frac{\rm N}{\rm m}$ \cite{grasigma} (the breaking strength can be estimated as the elastic moduli divided by $2\pi$ \cite{frenkel26,bukreeva}). This $\sigma$ is of the same order as the upper theoretical bound \eqref{sigma} and conforms to this bound.

We can also consider a one-dimensional chain and introduce a theoretical limit for the elastic force $f$ acting as one-dimensional analogue of the elastic modulus as:

\begin{equation}
f=\frac{E_{\rm R}}{a_{\rm B}}\approx 41~{\rm nN}
\label{force1}
\end{equation}

\noindent or

\begin{equation}
f\propto\left(\frac{m_ee^3}{\hbar^2}\right)^2
\label{force2}
\end{equation}

\noindent in terms of fundamental constants.

Similarly to using graphene to compare the theoretical prediction of two-dimensional elasticity in terms of funamental constants, we can look to compare $f$ in Eq. \eqref{force1} to one-dimensional structures of carbon, carbyne. Experiments in carbyne report $f=8-12$ nN \cite{carbyne1,carbyne2}. This is of the same order as the theoretical upper bound \eqref{force1} and conforms to this bound.

This review is largely related to three-dimensional condensed matter systems. It is nevertheless interesting to note that low-dimensional systems and their derivatives can offer a particularly simple relationship between system properties and fundamental physical constants. An interesting result comes again from graphene: the light absorption coefficient of a single graphene layer is theoretically predicted to be $\pi\alpha$, where $\alpha$ is the fine structure constant we discussed earlier \cite{katsnelson}, in agreement with experimental results \cite{geim}.

We also note that we do not consider effects directly related to quantization in this review. If a property changes in quanta (e.g., resistivity quanta in the quantum Hall effect or magnetic flux quanta in superconductors), bounds related to the smallest quantum number can trivially emerge too. The nature of these bounds is different from those considered in this review: we discuss bounds whose origin is unrelated to quantization.

\section{Speed of sound}
\label{sound}

\subsection{The upper bound}

Our last case study in this review is the speed of sound in condensed matter phases, $v$, and its upper limit in terms of fundamental physical constants.

There are two approaches in which $v$ can be evaluated. The first approach involves elasticity involving Eq. \eqref{elenergy3}.

The longitudinal speed of sound is $v=\left({\frac{M}{\rho}}\right)^{\frac{1}{2}}$, where $M=K_0+\frac{4}{3}G$, $K$ is the bulk modulus, $G$ is the shear modulus, and $\rho$ is the density. As discussed in Section \ref{elastic1}, the bulk modulus is governed by the density of cohesive energy in Eq. \eqref{elenergy3}, where the proportionality coefficient $f=f_0^{''}$ is experimentally found to be in the range 1-4 \cite{brahemley,solozhen}. The same data implies the proportionality coefficient between $M$ and $\frac{E}{a^3}$ in the range of about 1-6. Combining $v=\left({\frac{M}{\rho}}\right)^{\frac{1}{2}}$ and $M=f\frac{E_b}{a^3}$ gives $v=f^{\frac{1}{2}}\left(\frac{E}{m}\right)^{\frac{1}{2}}$, where $m$ is the mass of the atom or molecule, and we used $m=\rho a^3$. The factor $f^{\frac{1}{2}}$ is about 1-2 and can be dropped in an approximate evaluation of $v$. Then,

\begin{equation}
v=\left(\frac{E}{m}\right)^{\frac{1}{2}}
\label{v01}
\end{equation}

We now recall that the bonding energy in condensed phases is given by the Rydberg energy in Eq. \eqref{rydberg}. Using $E=E_{\rm R}$ from Eq. (\ref{rydberg}) in (\ref{v01}) gives

\begin{equation}
\frac{v}{c}=\alpha\left(\frac{m_e}{2m}\right)^{\frac{1}{2}}
\label{v00}
\end{equation}

\noindent where $\alpha=\frac{1}{4\pi\epsilon_0}\frac{e^2}{\hbar c}$ is the fine structure constant.

The second approach to evaluating the speed of sound involves the vibrational properties of the system. The longitudinal speed of sound, $v$, can be evaluated as the phase velocity in the longitudinal dispersion curve $\omega=\omega$($k$): $v=\frac{\omega_{\rm D}}{k_{\rm D}}$, where $\omega_{\rm D}$ and $k_{\rm D}$ are Debye frequency and wavevector, respectively. Using $k_{\rm D}=\frac{\pi}{a}$, where $a$ is the interatomic or inter-molecule separation, gives

\begin{equation}
v=\frac{1}{\pi}\omega_{\rm D}a
\label{v001}
\end{equation}

Using the ratio between the phonon energy, $\hbar\omega_{\rm D}$, and $E$ in Eq. \eqref{ratio} in Section \ref{vminima} in Eq. \eqref{v001} gives

\begin{equation}
v=\frac{Ea}{\pi\hbar}\left(\frac{m_e}{m}\right)^{\frac{1}{2}}
\label{v12}
\end{equation}

$v$ in (\ref{v00}), up to a constant factor, is obtained by using $a=a_{\rm B}$ from (\ref{bohr}) and $E=E_{\rm R}$ from (\ref{rydberg}) in (\ref{v12}). Alternatively, the same result can be found by using $E=\frac{\hbar^2}{2m_ea^2}$ \eqref{direct} and $a=a_{\rm B}$ (\ref{bohr}) in (\ref{v12}).

Compared to the first approach, the second approach based on vibrational properties involves additional approximations, including evaluating $v$ from the dispersion relation in the Debye model, using $a=a_{\rm B}$ in (\ref{bohr}) and the ratio between the phonon and bonding energies (\ref{ratio}). For this reason, we focus on the result from the first approach, Eq. (\ref{v00}).

In Eq. \eqref{v00}, $m_e$ characterises electrons, which are responsible for the interactions between atoms. The electronic contribution is further reflected in the factor $\alpha c$ ($\alpha c\propto\frac{e^2}{\hbar}$), which is the electron velocity in the Bohr model. We note that $v$ does not depend on $c$. The reason for writing the fraction $\frac{v}{c}$ in terms of $\alpha$ is two-fold. First, this ratio is convenient and informative, similarly to the ratio of the Fermi velocity and the speed of light $\frac{v_{\rm F}}{c}$ commonly used. Second, it is $\alpha$ (together with $\frac{m_e}{m_p}$) that is given fundamental importance and is finely tuned to enable the synthesis of heavy elements \cite{barrow} and, therefore, the existence of solids and liquids where sound can propagate to begin with.

Similarly to the viscosity minimum $\nu_{min}$ and fundamental viscosity $\nu_f$ in section \ref{vminima}, the derivation of Eq. \eqref{v00} involves more than dimensional analysis. First, the dimensional analysis alone is consistent with re-writing Eq. \eqref{v00} as $\frac{v}{c}=f_1(\alpha)f_2\left(\frac{m_e}{m}\right)$, where $f_1$ and $f_2$ are arbitrary functions. This gives any desired value for the speed of sound. Instead, the derivation of Eq. \eqref{v00} involves several physical insights not available in the dimensional analysis alone. This includes the experimental result of $f_0^{''}$ being close to 1 in Eq. \eqref{elenergy2}, using the ratio $\frac{\hbar\omega_{\rm D}}{E}$, based on the physical model, in Eq. \eqref{v12}, and so on. These steps are physically guided and incorporate more information that would be available from purely dimensional considerations.

$m$ in (\ref{v00}) characterises atoms involved in periodic motion during sound propagation. The scale of $m$ is set by the proton mass $m_p$: $m=Am_p$, where $A$ is the atomic mass. Recall that $a_{\rm B}$ in (\ref{bohr}) and $E_{\rm R}$ in (\ref{rydberg}) are characteristic values derived for the H atom. Setting $A=1$ and $m=m_p$ in (\ref{v00}) gives the upper bound of $v$ in (\ref{v00}), $v_u$, as \cite{sciadv2}:

\begin{equation}
v_u=\alpha\left(\frac{m_e}{2m_p}\right)^{\frac{1}{2}}c~\approx~36,100~\frac{\rm m}{\rm s}
\label{v36}
\end{equation}

We observe that $v_u$ depends on fundamental physical constants including the dimensionless fine structure constant $\alpha$ and the proton-to-electron mass ratio. We have discussed the importance of these two constants earlier in this review, including in the Introduction.

Combining Eqs. (\ref{v00}), (\ref{v36}), and $m=Am_p$ gives

\begin{equation}
v=\frac{v_u}{A^\frac{1}{2}}
\label{a}
\end{equation}

Before discussing Eq. (\ref{v00}) and its implications, Eqs. (\ref{v36})-(\ref{a}), we note that the speed of sound is governed by the elastic moduli and density which substantially vary with bonding type: from strong covalent, ionic, or metallic bonding, typically giving a large bonding energy to intermediate hydrogen-bonding, and weak dipole and van der Waals interactions. Elastic moduli and density also vary with the particular structure that a system adopts. Furthermore, structure and bonding type are themselves inter-dependent: covalent and ionic bonding result in open and close-packed structures, respectively \cite{phillips}. As a result, the speed of sound for a particular system can not be predicted analytically and without the explicit knowledge of structure and interactions. This is similar to other system-dependent properties such as viscosity or thermal conductivity discussed in Sections \ref{minimalv} and \ref{thermal} but is different to other properties such as the classical energy and specific heat which are universal in the harmonic approximation \cite{landaustat}. Nevertheless, the dependence of $v$ on $m$ or $A$ in Eq. \eqref{a} can be studied in a family of elemental solids. Elemental solids do not have confounding features of compounds related to mixed bonding between different atomic species, including mixed covalent-ionic bonding between the same atomic pairs as well as different bonding types between different pairs.

\subsection{Comparing to experiments}

\begin{figure}
{\scalebox{0.37}{\includegraphics{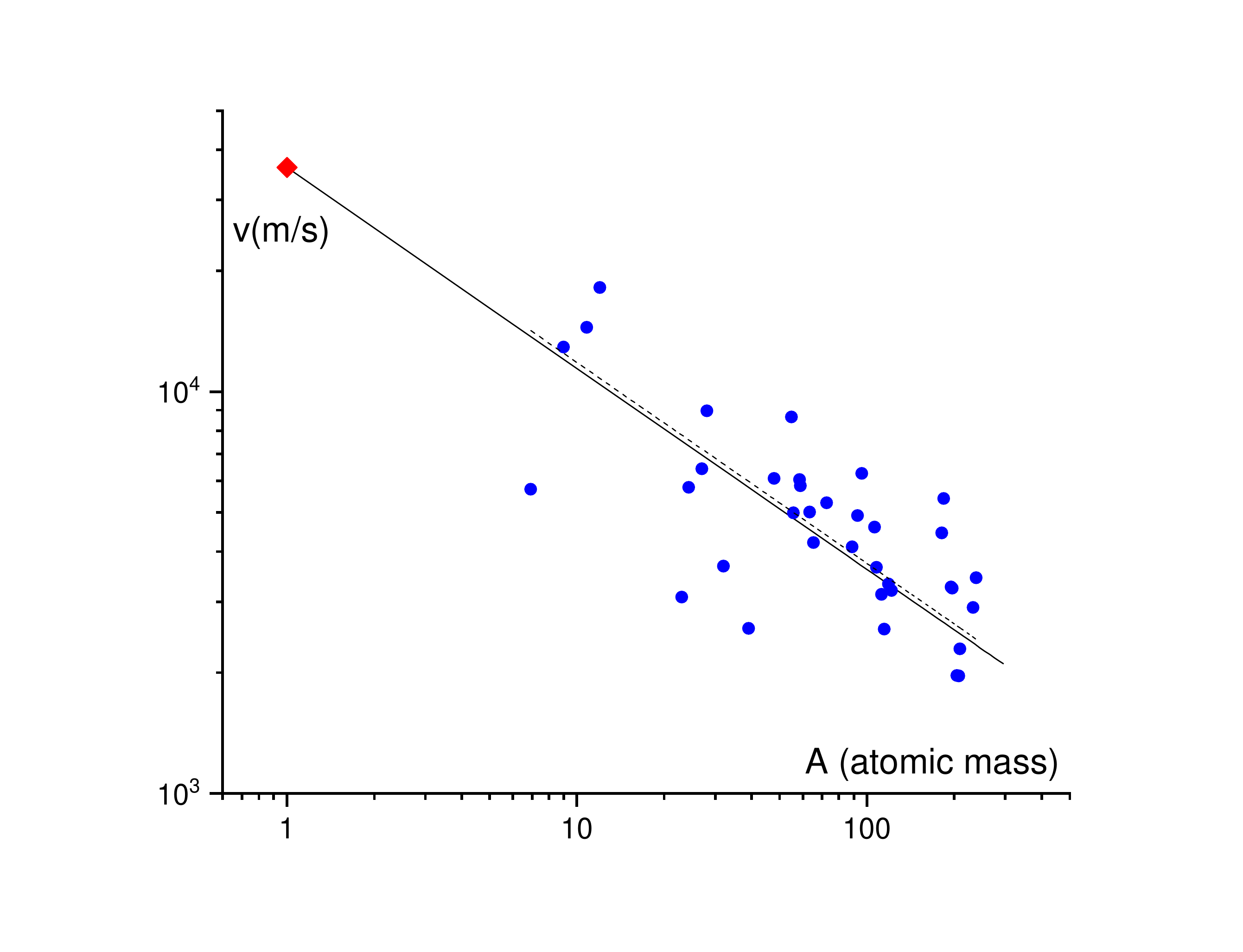}}}
\caption{Experimental longitudinal speed of sound \cite{handbook,handbook1,drits} in 36 elemental solids (blue bullets) as a function of atomic mass. The solid line is the plot of Eq. (\ref{a}): $v=\frac{v_u}{A^\frac{1}{2}}$. The red diamond shows the upper bound of the speed of sound (\ref{v36}). The dashed line is the fit to the experimental data points. In order of increasing mass, the solids are: Li, Be, B, C, Na, Mg, Al, Si, S, K, Ti, Mn, Fe, Ni, Co, Cu, Zn, Ge, Y, Nb, Mo, Pd, Ag, Cd, In, Sn, Sb, Ta, W, Pt, Au, Tl, Pb, Bi, Th and U. From Ref. \cite{sciadv2}.}
\label{elemental}
\end{figure}

The implication Eq. \eqref{v00} leading to the upper bound \eqref{v36} is Eq. \eqref{a}. We can compare Eq. \eqref{a} to experiments. We plot the available data of $v$ as a function of $A$ for 36 elemental solids \cite{handbook,handbook1,drits} in Fig. \ref{elemental}, including semiconductors and metals with large bonding energies. Eq. \eqref{a} is the straight line in Fig. \ref{elemental} ending in the upper theoretical bound (\ref{v36}) for $A=1$. The linear Pearson correlation coefficient calculated for the experimental data ($\log A$, $\log v$) is $-0.71$. Its absolute value is slightly above that notionally separating moderate and strong correlations \cite{correlation}. We also find that the ratio of calculated and experimental $v$ is in the range 0.6-2.4, consistent with the range of $f^\frac{1}{2}$ approximated by 1 in the derivation of Eq. (\ref{v01}).

The dashed line in Fig. \ref{elemental} shows the fit of the experimental data points to the inverse square root function predicted by Eq. (\ref{a}) and lies very close to Eq. (\ref{a}). The fitted curve gives the intercept at 37,350 $\frac{\rm m}{\rm s}$. This is in about 3\% agreement with the upper bound $v_u$ in (\ref{v36}). This indicates that the numerical coefficient in Eq. (\ref{v00}), which is subject to an approximation as mentioned earlier gives good agreement with the experimental trend. The agreement of Eq. (\ref{a}) with experimental data supports Eq. (\ref{v00}) and its consequence, the upper limit $v_u$ in Eq. (\ref{v36}).

\begin{figure}
{\scalebox{0.37}{\includegraphics{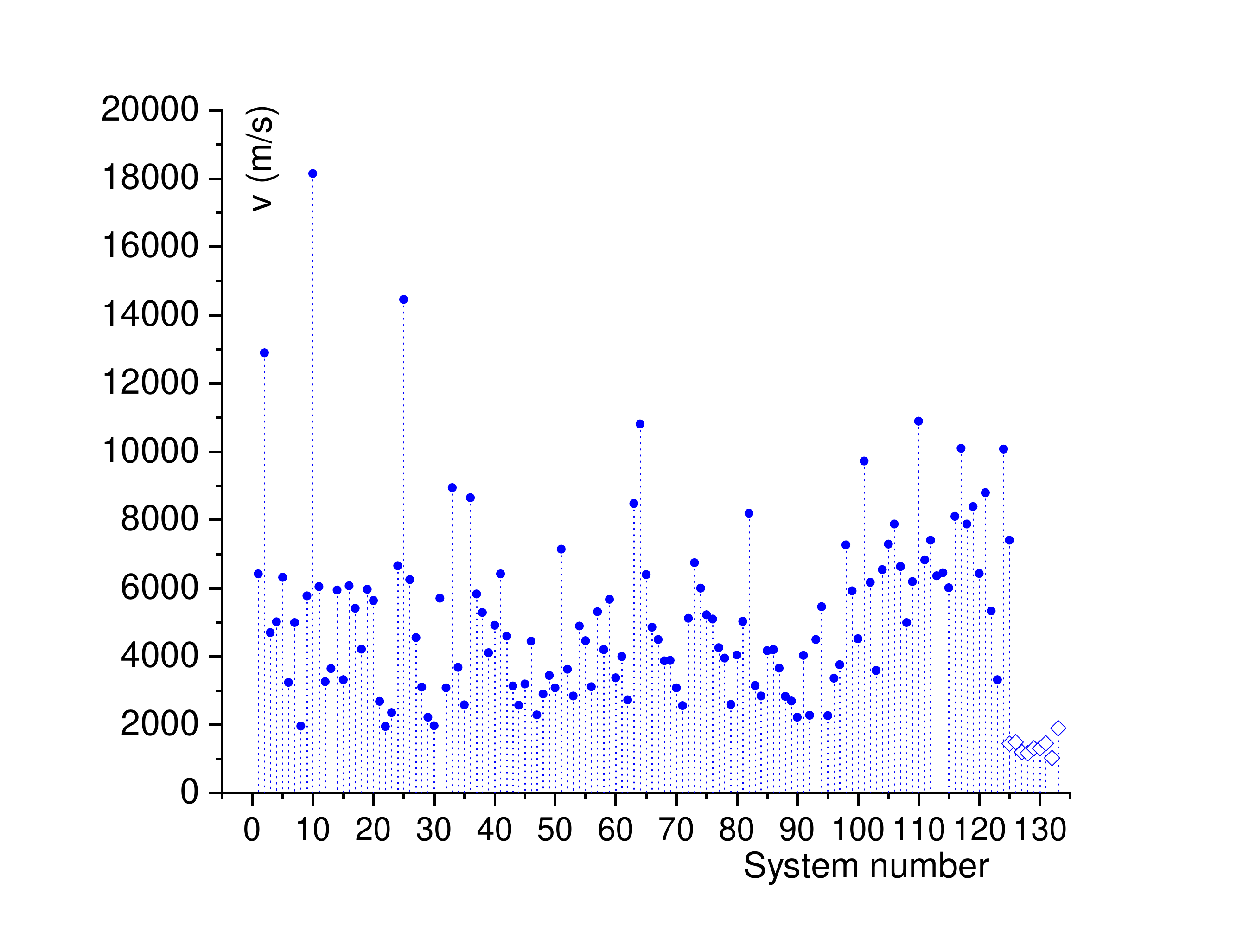}}}
\caption{Experimental longitudinal speed of sound \cite{handbook,handbook1,drits} in 124 solids (circles) and 9 liquids \cite{handbook} (diamonds) at ambient conditions as a function of the system number. Solids are: Al, Be, Brass, Cu, Duralumin, Au, Fe, Pb, Mg, Diamond, Ni, Pt, Ag, Steel, Sn, Ti, W, Zn, Fused silica, Pyrex glass, Lucite, Polyethylene, Polyesterene, WC, B, Mo, NaCl, RbCl, RbI, Tl, Li, Na, Si, S, K, Mn, Co, Ge, Y, Nb, Mo, Pd, Cd, In, Sb, Ta, Bi, Th, U, LiF, LiCl, BeO, NH$_4$H$_2$PO$_4$, NH$_4$Cl, NH$_4$Br, NaNO$_3$, NaClO$_3$, NaF, NaBr, NaBrO$_3$, NaI, Mg$_2$SiO$_4$, $\alpha$-Al$_2$O$_8$, AlPO$_4$, AlSb, KH$_2$PO$_4$, KAl(SO$_4$)$_2$, KCl, KBr, KI, CaBaTiO$_3$, CaF$_2$, ZnO, $\alpha$-ZnS, GaAs, GaSb, RbF, RbBr, Sr(NO$_3$)$_2$, SrSO$_4$, SrTiO$_3$, AgCl, AgBr, CdS, InSb, CsCl, CsBr, CsI, CsF, Ba(NO$_3$)$_2$, BaF$_2$, BaSO$_4$, BaTiO$_3$, TlCl, Pb(NO$_3$)$_2$, PbS, Apatite, Aragonite, Barite, Beryl, Biotite, Galena, Hematite, Garnet, Diopside, Calcite, Cancrinite, Alpha-quartz, Corundum, Labradorite, Magnetite, Microcline, Muscovite, Nepheline, Pyrite, Rutile, Staurolite, Tourmaline, Phlogopite, Chromite, Celestine, Zircon, Spinel and Aegirite. Liquids are: Mercury, Water, Acetone, Ethanol, Ethylene, Benzene, Nitrobenzene, Butane and Glycerol. See Refs. \cite{handbook,handbook1,drits} for system specifications, including density and symmetry groups. From Ref. \cite{sciadv2}.}
\label{all}
\end{figure}

We can also see that $v_u$ agrees with a wider experimental set. In Fig. \ref{all}, we show experimental $v$ \cite{handbook,handbook1,drits} in 133 systems, including elementals systems and compounds. As expected, the experimental $v$ are smaller than the upper theoretical bound $v_u$ in (\ref{v36}). $v_u$ is about twice as large as $v$ in diamond, the highest $v$ measured at ambient conditions (the in-plane speed of sound in graphite is slightly above $v$ in diamond \cite{behnia1}).

Eq. (\ref{a}) can be used to roughly predict the average, or characteristic speed of sound $v$ in condensed matter systems. $A^{\frac{1}{2}}$ which, according to (\ref{a}) is relevant for $v$, varies across the periodic table in the range of about 1-15, with an average value of 8. According to (\ref{a}), this corresponds to $v\approx 4,513\frac{\rm m}{\rm s}$. This is in 16\% agreement with 5,392 $\frac{\rm m}{\rm s}$, the average over elemental solids in Fig. \ref{elemental} and in 14\% agreement with 5,267 $\frac{\rm m}{\rm s}$, the average over the wider range of solids in Fig. \ref{all}.

This explains the characteristic values of $v$ and their average. Although $v$ depends on the system in \eqref{a}, the scale of $v$ is defined by Eq. \eqref{v36} which is set by fundamental physical constants. Earlier in this review, we have seen that characteristic values of other properties, including viscosity, thermal diffusivity and elasticity are similarly set by fundamental constants.

Fig. \ref{all} includes the experimental $v$ of room-temperature liquids with typical $v$ in the range 1,000-2,000 $\frac{\rm m}{\rm s}$. $v$ in high-temperature liquid metals such as Al, Fe, Mg, and Ni is in the higher range 4,000-5,000 $\frac{\rm m}{\rm s}$ \cite{metals}. We see that $v$ in liquids satisfy the bound $v_u$, similarly to solids. We note that the evaluation of $v$ and $v_u$ applies to liquids with cohesive states where molecular dynamics includes solid-like oscillatory components below the Frenkel line \cite{flreview} discussed in Section \ref{minimalv}. In this regime, $v$ is set by the elastic moduli as in solids albeit taken at their high-frequency (short-time) values \cite{frenkel} so the derivation in Eqs. \eqref{v01}-\eqref{v36} applies. On the other hand, at high temperature and/or low density above the Frenkel line, cohesive states are lost and Eq. (\ref{rydberg}) and Eq. (\ref{bohr}) and the derivation of $v$ do not apply. In this regime, the moduli are related to the kinetic energy of molecules rather than interactions and bonding energy, and $v$ starts to increase with temperature and loses its universality. Above the Frenkel line \cite{flreview} where the molecular motion is purely diffusive, $v$ is equal to the thermal speed of molecules as in a gas.

In Sections \ref{minimalv} and \ref{thermal}, we have discussed fundamental bounds of $\nu$ and $\alpha$ and later saw that they represent the bounds for all states of matter, including solids and gases. In this regard, it is interesting to note that an expression similar to (\ref{v01}) was earlier obtained by evaluating the elastic modulus using the liquid state theory and applied to liquid metals \cite{gitis}. The speed of sound was also evaluated in the theory of metals using the ionic plasma frequency and subsequently accounting for the conduction electrons screening. This results in the Bohm-Staver relation $v\propto\left(\frac{m_e}{m}\right)^{\frac{1}{2}}v_{\rm F}$, where $v_{\rm F}$ is the Fermi velocity \cite{ashcroft}, and hence $v\propto\frac{1}{A^\frac{1}{2}}$ as in Eq. (\ref{a}). These and other relations derived for the liquid state give a fairly good account of the experimental sound velocity in metallic liquids \cite{gitis,metals}.

We make three remarks about the calculated $v$ and its bound. First, this derivation involves approximations as mentioned earlier, which may affect the numerical factor in Eqs. (\ref{v00}) and (\ref{v36}). However, the characteristic scale of $v$ in (\ref{v00}) and its upper bound (\ref{v36}) is set by fundamental physical constants. The second remark is similar to the disclaimer we made with regard to liquids in Section \ref{minimalv}: Eq. (\ref{rydberg}) as well as Eqs. (\ref{v00})-(\ref{v12}) used in the second approach to derive $v$ assume valence electrons directly involved in bonding and hence strongly-bonded systems, including covalent, ionic and metallic systems. Although bonding in weakly-bonded solids such as noble, molecular and hydrogen-bonded solids is also electromagnetic in origin, weak dipole and van der Waals interactions result in smaller $E$ \cite{vadim1} and smaller $v$ as a result. Therefore, the upper bound $v_u$ in Eq. \eqref{v36} applies to weakly-bonded systems too.


The upper bound in Eq. (\ref{v36}) corresponds to solid hydrogen with strong metallic bonding. Although this phase only exists at megabar pressures \cite{silvera,loubeyre} and is dynamically unstable at ambient pressure where molecular formation occurs, $v$ can be calculated in atomic hydrogen using quantum-mechanical calculations. This carries an additional interest due to research into the properties of atomic hydrogen at high pressure (see, e.g., Refs. \cite{silvera,loubeyre,hydrogen}), although the speed of sound in these phases was not discussed and remains unknown. The quantum-mechanical calculation of $v$ shows good agreement with Eq. (\ref{v36}) \cite{sciadv2}.

We make several remarks related to previous work involving bounds on the speed of sound. It was noted that thermal diffusivity of insulators does not fall below a threshold value given by the product of $v^2$ and the Planckian time \cite{behnia}. Later work linked the upper bound on the speed of sound to the melting velocity related to melting temperature and Lindemann criterion \cite{hartnoll1}. In hadronic matter, the upper bound of the speed of sound was conjectured to be \cite{cherman,hadronic}):

\begin{equation}
v_u=\frac{c}{\sqrt{3}}
\label{hadronic}
\end{equation}

Later work \cite{breaking1,breaking2,breaking3} discussed the bound \eqref{hadronic} and its violations using different models. Comparing the bound \eqref{hadronic} with (\ref{v36}), we see that the bound \eqref{v36} is smaller due to small coupling constant $\alpha$ and the electron-to-proton mass ratio. In hadronic matter with strong coupling and particles with the same or similar masses, these factors become close to 1, in which case $\frac{v_u}{c}$ in Eq. (\ref{v36}) becomes closer to the conjectured limit.

We finally note that the upper bound for $v$ plays a role in thermodynamic properties too. Indeed, the low-temperature entropy and heat capacity per volume in solids are

\begin{eqnarray}
\begin{split}
& \frac{C}{V}=\frac{2\pi^2}{5(\hbar u)^3}T^3\\
& \frac{S}{V}=\frac{2\pi^2}{15(\hbar u)^3}T^3
\end{split}
\label{cs}
\end{eqnarray}

\noindent where $u$ is the average speed of sound \cite{landaustat}.

Hence, the upper bound for $u$ gives the smallest possible entropy and heat capacity at a given temperature.

\section{Summary: fundamental constants and physical theories}
\label{summary}

Testing and validating a physical theory and comparing it to an experiment involves different types of numbers, and in this sense Eq. \eqref{cs} serves as a good representative example. There are ordinary numbers such as ``2'' or ``$\pi$'' in Eq. \eqref{cs}. Then there is a family of external parameters such as pressure, temperature, external field and so on. In Eq. \eqref{cs}, this is represented by temperature $T$. Another type of parameters are fixed by system properties. In Eq. \eqref{cs}, this is the speed of sound $u$. Parameters like this are different from free adjustable parameters. We prefer not to have those in our theories if possible, not least because it may be hard to judge whether the theory is valid if parameters are freely adjustable. We can often fit experimental data with a fairly small number of adjustable parameters and hence can't decide which theory is the correct one, remaining no wiser as to what physical mechanism really operates. On the other hand, a parameter fixed by system properties has no such flexibility. If we know $u$ from some other experiment or simulation, Eq. \eqref{cs} unambiguously predict heat capacity and entropy at a given temperature. This is what physics is considered to be about: one view holds that the essence of every physical theory is to predict a future experiment on the basis of a previous one \cite{landaupeierls} or, in other words, provide a relationship between different properties.

Finally, there is another class of parameters in a theory: fundamental physical constants. In Eq. \eqref{cs}, this is $\hbar$. Being the ``barcodes of ultimate reality'' \cite{barrow}, these are very special parameters. They too are fixed by system properties as in the earlier example, with the proviso that the system is the Universe.

Accordingly, a theory where an observable is expressed in terms of fundamental constants only (as well as ordinary numbers) is a special type of theory because it directly links the property in question to the Universe properties.

By design, such a theory does not address effects related to variation of external parameters (pressure, temperature and so on). Applied to a range of systems, such a theory often makes approximations which is inevitable in view of structural, chemical and bonding variety of condensed matter phases. Once these are made, the theory and its results allow no further flexibility or leeway. With this disclaimer, we summarise what we can learn from such a theory below.

We have seen that fundamental physical constants can usefully provide a {\it bound} on a physical property. We have discussed how this works for properties which are quite complicated to be amenable to a general theoretical treatment including liquid viscosity and thermal conductivity.

We found that $\hbar$ interestingly enters all expressions for bounds on macroscopic properties. This includes the range of external parameters where systems are considered classical. Viscosity and thermal diffusivity bounds are additionally governed by $m_e$ and $m_p$.

We have seen that comparing the observed property to its fundamental bound informs us about the dynamical regime the system is in. For example, if viscosity or thermal diffusivity are close to its lower bound $\nu_{min}$ ($\alpha_{min}$), we are able to conclude how particle move, namely that they are close to the dynamical crossover between the liquidlike and gaslike motion. This contains quite a lot of information which is not at all easy to ascertain on the basis of experiments or even modelling, and yet we are able to make this assertion on the basis of one measured number only: $\nu_{min}$ or $\alpha_{min}$.

The value of bounds in terms of fundamental physical constants also provide a consistency check for a theory because it anchors the limiting values predicted by the theory. It also serves as a useful guide for a future theory, as illustrated by the discussion of quantum liquids and microscopic theory of liquid He which is yet to be developed.

Knowing the bounds from fundamental constants is also useful in systems where high-pressure and high-temperature properties have not been measured yet due to high melting points, such as molten salts or liquid metals.

From the practical standpoint, the fundamental bounds are useful in a number of ways. For example, they inform us that we should not spend an effort to design a lubricating fluid with viscosity significantly below the lower viscosity bound because this would not be allowed by the fundamental constants. On the other hand, if the measured values of viscosity or thermal conductivity are well above the bound, then there is a room for improvement that can be pursued.

In addition to setting the bounds, fundamental constants explain the observed characteristic values of several important properties of condensed matter in ways not anticipated before. Recall the calculated average value of the speed of sound in solids of about 5 km/s. By relating this value to fundamental constants, we can understand why it is 5, rather than, for example, 50 or 500 km/s. Similarly, we can understand why viscosity, thermal conductivity and elastic properties of many systems of interest take the values they do. This understanding is similarly provided by fundamental physical constants. One of the main points of this review was to show how this happens.

The observation that the fundamental physical constants largely govern water viscosity at ambient conditions raises more general and far-reaching questions related to our place in the Universe: what values of these constants make water-based life possible? What happens to water-based life forms if fundamental physical constants change, and how finely-tuned do these constants need to be to remain bio-friendly? This adds another, biochemical, layer to the discussion of the anthropic principle \cite{barrow,barrow1,hoganreview,adamsreview,uzanreview} and invites an inter-disciplinary research.

Some of the fundamental bounds we discussed in this review were known, whereas others are fairly new and unexpected. There seems to have been no sustained work in this area, and this review may encourage further thinking and discovery of new fundamental bounds.

\section{Acknowledgements}

I am grateful to the EPSRC for support, to V. V. Brazhkin, B. Carr, L. Noirez, G. Volovik and U. Windberger for discussions and to V. V. Brazhkin for contributing to Section \ref{lowd} in 2021.

%

\bibliographystyle{apsrev4-1}

\end{document}